\documentclass[a4paper,10pt]{article}
\usepackage{times}
\usepackage{graphicx} 
\usepackage{bmpsize}
\usepackage{amsfonts}
\usepackage{amsmath}
\usepackage{amssymb}
\usepackage{amstext}
\usepackage{amsthm}
\usepackage{diagram}
\usepackage[all]{xy}
\usepackage[utf8]{inputenc}
\usepackage{microtype}
\usepackage{amssymb}
\usepackage{mathtools}
\usepackage{bbold}
\usepackage{tikz}
\usetikzlibrary{matrix,arrows,trees}
\usepackage{tikz-cd}
\usepackage{hyperref}
\usepackage[capitalize, nameinlink]{cleveref}
\usepackage{todonotes}
\usepackage{xcolor}

\newcommand{\be}{\begin{equation}}
\newcommand{\ee}{\end{equation}}
\newcommand{\bea}{\begin{eqnarray}}
\newcommand{\eea}{\end{eqnarray}}
\newcommand{\ba}{\begin{array}}
\newcommand{\ea}{\end{array}}

\newcommand{\la}{\label}

\makeatletter \@addtoreset{equation}{section} \makeatother

\begin{document}

\begin{titlepage}

    \thispagestyle{empty}
  %  \begin{flushright}
      %  \hfill{SP-ITP-14/...}\\
    %\end{flushright}

    \vspace{82pt}
    \begin{center}
        { \Huge{\bf Towards a T-dual Emergent Gravity}}

        \vspace{18pt}
        {\large{\bf Daniel Bermudez $^{\heartsuit}$ and \bf Raju Roychowdhury$^{\clubsuit, \diamondsuit}$}}

        \vspace{15pt}

        {$\heartsuit$ \it
 Mathematical Institute, University of Bonn,\\
                53012, Bonn, Germany\\
        \texttt{danielbermudez@uni-bonn.de}}

\vspace{10pt}

        {$\clubsuit$  \it Dept. of Physics, AppLied Sciences Cluster\\
School of Advanced Engineering\\
UPES, Dehradun- 248007, India}

\vspace{10pt}        

 {$\diamondsuit$ \it
 Department of Physics and Astrophysics\\
 University of Delhi,
Delhi 110 007, India\\
        \texttt{rroychowdhury@physics.du.ac.in}}

       \end{center}       

\vspace{85pt}

\begin{abstract}
Emergent gravity provides a geometric interpretation of noncommutative $U(1)$ gauge theory, in which deformations of a symplectic structure induced by a gauge field are absorbed by diffeomorphisms via the Darboux theorem, giving rise to an effective Riemannian metric. Independently, topological T-duality identifies pairs of principal torus bundles equipped with distinct geometric and flux data whose associated sigma models are physically equivalent. In this work we place both constructions within the framework of generalized geometry, thereby enabling a uniform treatment of emergent metrics and T-duality transformations. We describe emergent gravity in terms of generalized metrics on exact Courant algebroids and interpret the Seiberg–Witten map as a composition of $B$ and $\theta$ transformations acting on a flat background. Using the \emph{Gualtieri–Cavalcanti formulation} of T-duality as an isomorphism of Courant algebroids, we construct a natural notion of a T-dual emergent gravity for principal torus bundles. For flat spacetimes with trivial H-flux, we show that the T-dual generalized metric again admits an emergent gravity interpretation. At the level of generalized geometry, this result is encoded in a commutative diagram in which T-duality interchanges the order of the transformations responsible for the emergence of the metric. For general $\mathbb{T}^2$-fibrations we derive explicit formulae for the T-dual generalized metric and show that the dual background generically carries a nontrivial H-flux which obstructs a direct interpretation in terms of conventional symplectic emergent gravity and necessitates an extension of the framework to non-exact Courant algebroids. Our results provide a precise mathematical link between emergent gravity and T-duality and indicate that generalized geometry furnishes the natural setting for a T-duality–covariant formulation of emergent gravity, including backgrounds with nonvanishing flux.

\end{abstract}

%\tableofcontents

\end{titlepage}

\section{Introduction}
Gauge theories on noncommutative (NC) spacetime are significantly better understood than gravity formulated in the same setting. The Seiberg–Witten (SW) map \cite{ncft-sw} provides a crucial tool for translating a gauge theory defined on a flat NC space into an interacting non-linear gauge theory on a curved background, where the background geometry itself is generated by the gauge field. When the gauge field possesses nontrivial curvature, a flat NC space can effectively give rise to a background with non-zero curvature. This phenomenon is interpreted as the emergence of a geometric structure from gauge theory on NC space and is often referred to as emergent gravity \cite{hsy-emergent,hsy-emergent2}. For a comprehensive review of emergent gravity and its quantization program, see the works \cite{lee1, lee2, lee3, hsy-quant} of Yang and collaborators.

\bigskip

More recently, a research group in Japan has developed a series of studies \cite{japan1, japan2, japan3} describing D-branes within the framework of generalized geometry. In their approach, D-branes-expressed in static gauge, including fluctuations-are identified with leaves of a foliation determined by the Dirac structure of the generalized tangent bundle. Scalar and gauge fields on the D-brane are treated uniformly as components of a generalized connection. In parallel, Jurčo et al. \cite{czech,czech2} have argued that the SW equivalence between commutative and semiclassically NC Dirac–Born–Infeld (DBI) actions is deeply rooted in the generalized geometric description of D-branes. More precisely, viewing the D-brane as a symplectic leaf of the Poisson structure encoding the noncommutativity, the SW map admits a natural interpretation in terms of the corresponding Dirac structure. This perspective positions generalized geometry as a natural and powerful framework for analyzing NC gauge theories. Furthermore, Yang et al. \cite{hsy-emergent, hsy-emergent2, lee1, lee2, lee3}  have provided arguments for why electromagnetism on NC spacetime should be regarded as a theory of gravity, demonstrating that NC electromagnetism \cite{paolo} can be realized via the Darboux theorem \cite{darboux} in symplectic geometry. This realization lies at the heart of emergent gravity, as it connects the deformation of a symplectic structure to diffeomorphism symmetry. These observations collectively suggest that the mathematical foundations of emergent gravity, intimately linked to NC electromagnetism, may be more fully understood through the formal tools offered by generalized geometry.

\bigskip

Generalized geometry, introduced by Hitchin \cite{gcg-hitchin} in the early 2000s and subsequently developed by Gualtieri and Cavalcanti \cite{gcg-gualtieri}, provides a unifying framework for the study of geometric structures on exact Courant algebroids. These algebroids can be viewed as the ``doubles" of the tangent bundle, given by the direct sum
$T \oplus T^*$ of the tangent and cotangent bundle of some smooth manifold $M$.  Beyond this purely algebraic construction, additional structures naturally arise, particularly those associated with a closed 3-form $H \in \Omega^3(M)$ satisfying $dH=0$.
This H-field plays a central role in string theory, where it corresponds to the Neveu–Schwarz (NS) flux, a closed 3-form appearing in the NS sector of type II supergravity backgrounds. Within this setting, the exact Courant algebroid furnishes a precise mathematical framework for describing the NS sector in terms of generalized geometry. The group $O(n, n ; \mathbb{R})$
acts naturally on the pairs
$(g,B)$, where $g$ is a Lorentzian metric on $M$ and
$B \in \Omega^{2}(M)$ is the Kalb-Ramond 2-form. This action is physically motivated: when restricted to the discrete subgroup $O(n, n ; \mathbb{Z})$, it identifies string backgrounds that are physically equivalent, a correspondence known as T-duality. The action $O(n, n ; \mathbb{R})$ is elegantly captured by the notion of a generalized metric $\mathcal{G}$ that encodes the pair
$(g,B)$ in a single geometric object. In his doctoral thesis \cite{gcg-gualtieri}, Gualtieri characterized type II string theory backgrounds preserving extended 
$\mathcal{N}=2$ supersymmetry as generalized Kähler manifolds. Later, together with Cavalcanti, he reformulated the T-duality as an isomorphism of Courant algebroids \cite{guacav}, thus deepening the geometric understanding of duality symmetries in string theory.

\bigskip

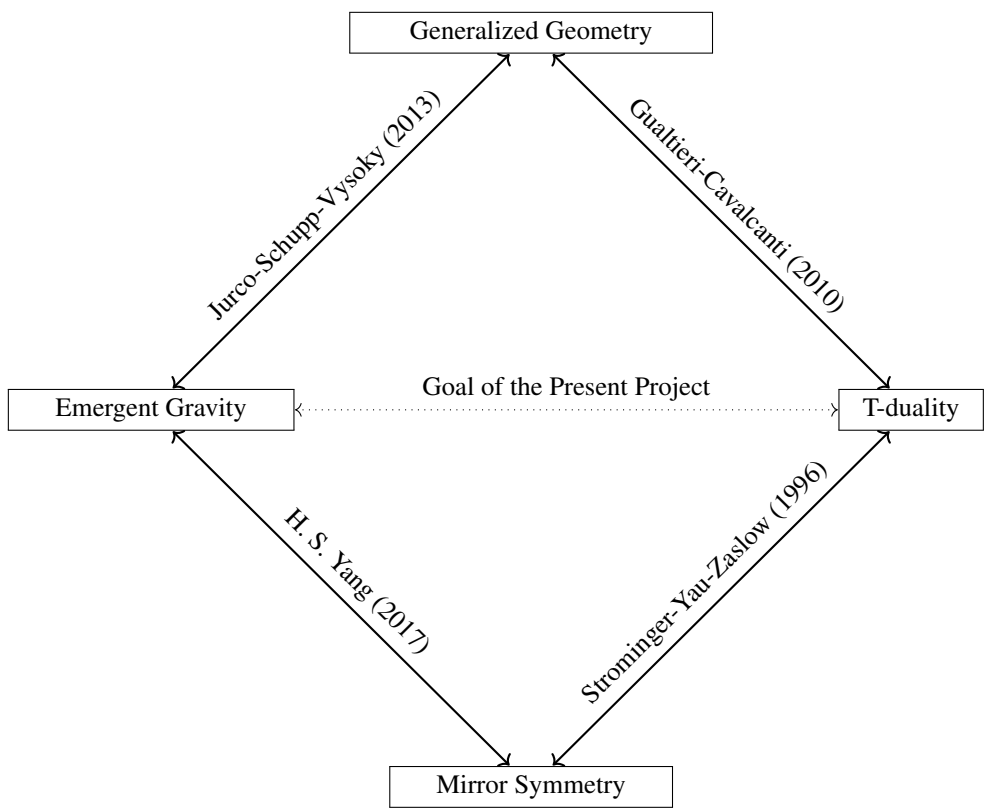
\begin{figure}
\label{project-block}
\centering

\begin{tikzpicture}
\node[inner sep=0pt] (Emergent) at (-5,0)
    {\framebox[1.5\width]{Emergent Gravity}};
\node[inner sep=0pt] (Generalized) at (0,5)
    {\framebox[1.5\width]{Generalized Geometry}};
\node[inner sep=0pt] (Tduality) at (5,0)
    {\framebox[1.5\width]{T-duality}};
    
\draw[<->,thick] (Generalized)--(Tduality) node[midway,sloped,above] {Gualtieri-Cavalcanti (2010)};
\draw[<->,thick] (Generalized)--(Emergent)node[midway,sloped,above] {Jurco-Schupp-Vysoky (2013)} ;
\draw[<->,dotted] (Emergent)--(Tduality)node[midway,sloped,above] {Goal of the Present Project} ;
\end{tikzpicture}

\caption{Flowchart for emergent gravity/T-duality correspondence.}
\label{fchart:emg}
\end{figure}

This project is motivated by an intuition summarized schematically in the triangular block diagram (\ref{project-block}). The guiding idea arises from three independent but closely related developments at the interface of gauge theory, geometry, and duality. First, Jurčo, Schupp, and Vysoký \cite{czech,czech2} showed that the Seiberg–Witten map establishing an equivalence between commutative and semi-classical noncommutative Dirac-Born-Infeld actions central to emergent gravity admits a natural and conceptually transparent formulation within generalized geometry. Second, the work of Gualtieri and Cavalcanti \cite{gcg-gualtieri, guacav} clarified the geometric content of topological T-duality by formulating it as an isomorphism of Courant algebroids, thereby providing a systematic framework for transporting generalized geometric structures between T-dual backgrounds. Third, developments in emergent gravity \cite{hsy-mirror} have emphasized that spacetime geometry itself can be viewed as an effective structure induced by deformations of an underlying symplectic or Poisson geometry associated with noncommutative gauge fields. Taken together, these three perspectives, represented as the vertices of the triangle in (\ref{project-block}), suggest that emergent gravity and T-duality should not be regarded as independent mechanisms, but rather as complementary manifestations of a common generalized geometric framework. In particular, they motivate the existence of a T-dual counterpart of emergent gravity, obtained by transporting its defining geometric data across T-dual pairs. The purpose of the present work is to make this relation explicit by identifying and analyzing the missing link, indicated by the dotted edge in (\ref{project-block}), between emergent gravity and T-duality.

\bigskip

The present work has several aims. First, we investigate T-duality in the framework of emergent gravity and elucidate its origin within generalized geometry. After identifying the generalized metric associated with $\mathbb{T}$-invariant emergent gravity, we employ the formalism developed by Gualtieri and Cavalcanti (GC) \cite{guacav} to transport both the generalized metric and the symplectic structure to the dual torus bundle. Our goal is to derive an explicit expression for the T-dual generalized structures obtained via Courant algebroid isomorphisms. For the case of a $\mathbb{T}^2$-action, we find that T-duality maps a symplectic structure to another symplectic structure, whereas in the case of an 
$\mathbb{S}^1$-action \cite{guacav}, T-duality relates a symplectic geometry on one side to a complex geometry on the other.

The main novel contributions of this work can be summarized as follows.
\begin{itemize}
    \item We derive explicit expressions for the T-dual of a symplectic form, leading to transformation laws that closely parallel the Buscher rules for the metric. These formulas, presented at the end of Section 4.3, provide a symplectic counterpart to the standard metric duality relations.

    \item We formulate emergent gravity within the language of generalized geometry and show that, in this framework, the Seiberg–Witten equations arise in a natural and conceptually transparent manner. This reinterpretation, developed in Section 5, clarifies the geometric origin of the open–closed string map.

    \item We obtain a genuine generalization of the Buscher rules to the case of metrics on $\mathbb{T}^{2}$ fibrations, whereas the conventional formulation applies primarily to $\mathbb{T}^{1}$ fibrations. The construction, detailed in the Appendix, is consistent with and complements the recent results of Waldorf et al.~\cite{WaldorfBuscher}, thereby providing an independent derivation from a generalized geometric perspective.
\end{itemize}

\bigskip

The paper is organized as follows. We begin with a concise glossary of key concepts related to emergent gravity, accompanied by a diagram intended to serve as a reference for subsequent sections. Section 2 introduces the basic framework of generalized geometry, leading to the notion of exact Courant algebroids (Section 2.1) in the context of emergent gravity. We consider examples of Courant algebroids associated with principal bundles and recall essential definitions, including the {\it Atiyah sequence} corresponding to the principal bundle underlying emergent gravity. Section 3 is devoted to the study of {\it generalized metrics}, understood here as generalizations of the tangent bundle metric on the manifold. Section 4 revisits the mathematical notion of T-duality, as formulated by Bouwknegt, Evslin, Mathai, and Hannabuss \cite{TopTdual1,TopTdual2, TopTdual3}, interpreting it as a correspondence between principal torus bundles. In Section 4.2, we propose a natural T-dual candidate for emergent gravity, formulated via the GC map, and provide a schematic diagram of topological T-duality for torus actions. Section 4.3 presents an explicit expression for the T-dual of the symplectic form with symplectic fibers, using the transport of invariant geometric structures under the GC map. Section 5 highlights the interplay between emergent gravity and T-duality, first conceptually and then through explicit computations for flat and curved spacetimes. In particular, we show that emergent gravity can be obtained by performing a B-transform followed by a 
$\theta$-transform on the flat metric, while in the T-dual description, the order of these operations is reversed, as depicted in a commutative diagram. Although the picture of emergent gravity in the presence of H-flux remains incomplete \footnote{The failure to obtain a T-dual of emergent gravity in the presence of H-flux has been investigated by one of the authors in \cite{RS}.}, we provide original formulas for the components of the emergent generalized metric and its T-dual counterpart, thereby laying the groundwork for a generalized geometric formulation of T-dual emergent gravity. Finally, Section 6 summarizes our results and outlines directions for future research. Additional computational details for Section 5.3 are provided in the Appendix.\\

\textbf{Glossary:}\\

\[\begin{tikzcd}
\mathbb{S}^1\circlearrowleft& (L,A)\ar[dd,"q"] & & (\hat{L},\hat{A})\ar[dd,"\hat{q}",dotted]\\
& & (M\times_B\hat{M},K) \ar[dl,"p"]\ar[dr,"\hat{p}",swap] & \\
\mathbb{T}^n\circlearrowleft& (M,b,F,H)\ar[dr,"\pi",swap] & & (\hat{M},\hat{b},\hat{F},\hat{H})\ar[dl,"\hat{\pi}"] & \\
& & B & &
\end{tikzcd}\]

\textbf{Emergent gravity data:}
\begin{itemize}
    \item $M$: The spacetime manifold (one that admits a symplectic structure).
    \item $b$: Symplectic structure on $M$  i.e. a non-degenerate closed 2-form.
    \item $L$: Line bundle over $M$-- $U(1)$-theory on $M$. 
    \item $A$: $U(1)$-connection of $L$-- potential of the $U(1)$-theory in $M$.
    \item $F$: The curvature of the connection $L$, that is, $F = dA$ - the field strength of the $U(1)$-theory.
    \item $\theta=b^{-1}$: The Poisson structure associated to $b$.
    \item $g$: The effective Riemmanian metric on $M$ determined by $b$ and $F$, to be precise, $g=g^{fl}(1+F\theta)$, where $g^{fl}$ is some flat metric on $M$. Since $g$ emerges from $U(1)$ gauge fields,
    we call it {\it emergent metric}.\footnote{In general $g$ might not be symmetric, because $g-g^{T} \propto (F\theta- \theta F) \neq 0$.    As is usual in the emergent gravity literature, see for example \cite{hsy-emergent2, lee1, lee2, lee3}, we will assume through the present work that $F\theta=\theta F$.} 
    \item  $\mathcal{G}$: The generalized metric on $TM \oplus T^*M$ determined by the pair $(g,b)$
        $$
\mathcal{G} = \left(
         \begin{array}{cc}
          - g^{-1}b & g^{-1} \\
           g- b g^{-1}b  & b g^{-1}  \\
         \end{array}
       \right),
    $$
    where $g$ is a Riemannian metric on $M$ and $b$ is a (symplectic) 2-form.%. $\mathcal{G}$ is the endomorphism of $TM \oplus T^*M$ defined :

    \item  $E_+ = \operatorname{graph}(g+b:TM \to T^*M)$: The graph associated to the generalized metric $\mathcal{G}$.
    
    \item $\pi\colon M\to B$ and $q\colon L\to M$: The projection maps that define the bundles $\mathbb{T}$ and $U(1)$, respectively.%; $p\colon M\times_B \hat{M}\to M$ is the natural projection map from the pullback to $M$. 
\end{itemize}

\textbf{T-dual emergent gravity data:}
\begin{itemize}
 \item $\hat{M}$: The dual spacetime.
     \item  $\mathcal{\hat{G}}$: The transformed generalized metric on $T\hat{M} \oplus T^*\hat{M}$ given by $\hat{E_+}$.
     \item $\hat{E}_+= \operatorname{graph}(\hat{g}+\hat{b}\colon T\hat{M} \to T^*\hat{M}$) where $(\hat{g},\hat{b})$ are given by the Buscher rules.
 \item $(a_i)$:  $\mathfrak{t}^n$-valued connection of the torus bundle $M\to B$: is a tuple of 1-forms, it gives a decomposition $T^*M=T^*B\oplus \langle a_1,a_2,\ldots,a_n \rangle$.
 \item $H$: The 3-form or the H-flux on $M$, in emergent gravity $H=db=0$, although $\hat{H}$ is not necessarily zero.
 \item $K$: The 2-form on the correspondence space such that $p^*H-\hat{p}^*\hat{H}=dK$. In the $H=0$ case, it is given by $K=\sum p^*a_i\wedge \hat{p}^*\hat{a}_i$. 
  \item {$\hat{L}$}: $U(1)$-bundle over $\hat{M}$ giving rise to a `T-dual' emergent gravity. It might not exist and conditions for its existence (in a special example) are discussed later in the work.
 \item $\hat{\pi}\colon\hat{M}\to B$ and $\hat{q}\colon\hat{L}\to \hat{M}$: The projection maps defined by the $\mathbb{T}$ and $U(1)$-bundle structures respectively (in case the latter exists).
 \item $\hat{p}\colon M\times_B \hat{M}\to \hat{M}$: The natural projection map from the pullback to $\hat{M}$. 
\end{itemize}

%\textbf{Note:} Everything with hat is the T-dual of the corresponding structure. We are yet to prove the existence of $(\hat{L},\hat{A})$ {\it\textbf{This is crucial for the T-dual gravity avatar to exist.}}

\section{Generalized Geometry and Emergent Gravity}

In classical differential geometry, a Riemannian structure is specified by a pair $(M,g)$ where $M$ is a smooth manifold and $g$ is a Riemannian metric
that encodes the full geometric information of the space.
Similarly, a symplectic structure is given by a pair $(M, b)$ where $b$ is a closed, non-degenerate 2-form providing the fundamental symplectic data.
Generalized geometry, introduced by Hitchin in 2002 \cite{gcg-hitchin} and subsequently developed by Gualtieri and Cavalcanti \cite{gcg-gualtieri}, offers a unifying framework that simultaneously incorporates both Riemannian and symplectic structures. This formalism treats the sum of the tangent and cotangent bundles, as the basic object of study, allowing metric and 2-form data to be described within a single geometric package. Central objects in this construction are the {\it Courant algebroids} 
which are the ambient spaces where one can define
geometric structures. Usually geometric structures are defined on manifolds but any manifold has a canonical
Courant algebroid associated to it and one can abstract that to do something more general.

\smallskip

Let $M$ be a smooth real manifold of dimension $n$ and $TM$ its tangent bundle.
To motivate the notion of  Courant algebroids let us consider the bundle 
$TM \oplus T^*M$  on $M$. The bundle $TM \oplus T^*M$ has a canonical $(n,n)$
signature pseudo-metric (a canonical fiberwise non-degenerate bilinear form)

\be \la{ip}
< X + \xi, Y + \eta > = \iota_X\eta + \iota_Y\xi = \eta (X) + \xi (Y),
\ee
where $X, Y \in \mathfrak{X}(M)$ are vector fields and $\xi, \eta \in \Omega^{1}(M)$ are 1-forms.  
\smallskip

We also have a natural projection $\pi \colon  E = TM \oplus T^*M \to TM$ to the first coordinate, which is a smooth bundle map called the anchor.
In the case of a differentiable manifold the space of sections of the tangent bundle is endowed with a
derived bracket, the Lie bracket, associated to the de Rham complex. It turns out that there is a derived bracket on the double bundle $TM \oplus T^*M$ which is mapped via the anchor $\pi$ to the Lie bracket. A skew-symmetric bracket on $TM \oplus T^*M$
was first introduced by Courant \cite{Dirac_Courant}, and later a non–skew-symmetric variant was formulated by Dorfman \cite{dorfman1987dirac}. Although differing in symmetry properties, the Courant and Dorfman brackets encode equivalent geometric data and are often used interchangeably in the literature. In this work, we adopt the convention of \cite{KS}, deriving the Dorfman (non–skew-symmetric) bracket in direct analogy with the construction of the Lie bracket of vector fields, by considering its action on differential forms.

\smallskip

On a smooth manifold $M$ vector fields act on the graded exterior algebra of
differential forms via the interior product
$X \cdot \varphi= \iota_X \varphi$, where $X\in \mathfrak{X}(M)$ and $ \varphi \in \Omega^{\bullet}(M)$.
The interior product is a degree-1 operator. We also have the exterior
differential $d$ acting on $\Omega^{\bullet}(M)$ which is of degree +1. The
Lie bracket of two vector fields $X$ and $Y$ is then
defined as the unique vector field satisfying
\be
[X, Y] \cdot \varphi=\left[\mathcal{L}_{X}, \iota_{Y}\right] \cdot \varphi=\left[\left[\iota_{X}, d\right], \iota_{Y}\right] \cdot \varphi 
\quad \text{ for all }  \varphi \in \Omega^{\bullet}(M) .
\ee
The commutators are meant to be supercommutators of operators 
\be
[A, B]=A \circ B-(-1)^{|A| \cdot|B|} B \circ A,
\ee
where $|A|$ denotes the degree of the operator $A$.

Notice that $TM \oplus T^*M$ also acts on forms via the Clifford action 
\be
(X+\xi) \cdot \varphi=\iota_{X} \varphi+\xi \wedge \varphi.
\ee
One can then define the Dorfman bracket of two sections $e_1, e_2 \in \Gamma(TM \oplus T^*M)$ as the unique section satisfying
\be
[e_1 , e_2]\cdot \varphi = [e_1, [d , e_2]] \cdot \varphi.
\ee
Although the elements of $TM \oplus T^*M$ have mixed tensorial type, both the vector and covector components are of odd degree in the graded sense. Consequently, $[d , e_1] = d \circ  e_1 + e_1 \circ d$.
Writing out the action one finds what is known as Dorfman bracket 
\be \la{dorfman}
[X + \xi, Y + \eta] = [X,Y] + {\cal L}_X \eta- \iota_Y d\xi 
\ee
of two sections $(X + \xi)$ and $(Y + \eta)$ on the space of sections of  $\Gamma(E)$.

\smallskip

We will usually denote by $\mathcal{E}$  the 4-tuple $(TM \oplus T^*M,\langle\cdot, \cdot\rangle, [\cdot,\cdot], \pi)$. This is the
first example of a {\it Courant algebroid}, objects that were axiomatized by
Liu, Weinstein and Xu in \cite{liu1997manin} for the skew-symmetrized version of
the bracket. We will not fully define Courant algebroids, since we are only interested in the case explained above; the interested reader is referred to \cite{liu1997manin} or \cite{guacav}. We will sometimes, by abuse of notation, denote a {\it Courant algebroid} $\mathcal{E}$ simply by $E$. We say that a Courant algebroid $\mathcal{E}$ is {\it transitive}  if the anchor $\pi \colon  E \to TM$ is surjective and that it is {\it exact} if it fits into the short exact sequence of vector bundles
\be
\la{exact}
0 \longrightarrow T^{*} M \stackrel{\frac{1}{2} \pi^{*}}{\longrightarrow} E \stackrel{\pi}{\longrightarrow} T M \longrightarrow 0.
\ee
Clearly $TM \oplus T^*M$ is an exact Courant algebroid over $M$.

%The following definition has been reformulated in terms of the nonskew-symmetric Dorfman bracket.
%{\bf Definition 2.0.1.} A Courant algebroid $\mathcal{E}$ over a manifold $M$ is a 4-tuple $(E,\langle\cdot, \cdot\rangle,[\cdot, \cdot], \pi)$ where $E \rightarrow M$ is a vector bunlde, $\langle\cdot, \cdot\rangle$ is a fibrewise non-degenerate symmetric bilinear pairing, $[\cdot, \cdot]$ is a bilinear bracket on the smooth sections $\Gamma(E)$ and $\pi: E \rightarrow T M$ is a smooth bundle map called the anchor such that the following axioms hold for all $e_{1}, e_{2}, e_{3} \in \Gamma(E)$ and $f \in \mathcal{C}^{\infty}(M)$ :
%\bea
%&Prop.(1)& \left[e_{1},\left[e_{2}, e_{3}\right]\right]=\left[\left[e_{1}, e_{2}\right], e_{3}\right]+\left[e_{2},\left[e_{1}, e_{3}\right]\right]\nonumber\\
%&Prop.(2)& \pi\left(\left[e_{1}, e_{2}\right]\right)=\left[\pi\left(e_{1}\right), \pi\left(e_{2}\right)\right]\nonumber\\
%&Prop.(3)& \left[e_{1}, f e_{2}\right]=\pi\left(e_{1}\right)(f) e_{2}+f\left[e_{1}, e_{2}\right]\nonumber\\
%&Prop.(4)& \pi\left(e_{1}\right)\left\langle e_{2}, e_{3}\right\rangle=\left\langle\left[e_{1}, e_{2}\right], e_{3}\right\rangle+\left\langle e_{2},\left[e_{1}, e_{3}\right]\right\rangle\nonumber\\
%&Prop.(5)& \left[e_{1}, e_{1}\right]=\mathcal{D}\left\langle e_{1}, e_{1}\right\rangle
%\eea
%where the bracket in $Prop.(3)$ on the right hand side is the Lie bracket of vector fields and in $Prop.(5)$ $\mathcal{D}=\frac{1}{2} \pi^{*} \circ d: \mathcal{C}^{\infty}(M) \rightarrow \Gamma(E)$ (using $\langle\cdot, \cdot\rangle$ to identify $E$ and $E^{*}$ ).
\subsection{Exact Courant Algebroids}

 We will now recall some of the geometry of Exact Courant algebroids, to see that the B-field featuring in emergent gravity has an interpretation as a Courant algebrioid automorphism. Exact Courant algebroids are slight generalizations of the Courant algebroid $TM \oplus T^*M$, indeed an isotropic splitting $s \colon  TM \to E$ of the sequence (\ref{exact}) gives a bundle isomorphism of $E$ with $TM \oplus T^*M$ which identifies the pairing on $E$ with the natural pairing in $TM\oplus T^*M$ . Whenever we talk about a splitting of $E$, we mean an isotropic splitting of (\ref{exact}). Isotropic splittings always exist as the inner product has split signature since the image of $T^*M$ is isotropic. Such a splitting $s \colon  TM \to E$ not only defines an isomorphism $E \cong TM \oplus T^*M$ but also a closed 3-form $H \in \Omega^3(M)$ via

\be
H(X, Y, Z) = \langle[s(X), s(Y)], s(Z)\rangle \quad \text{ for all }  X, Y, Z \in \mathfrak{X}(M).
\ee
\smallskip

Different splittings of $E$ are globally related by 2-forms $b \in
\Omega^2(M)$ which change $H$ by an exact 3-form. Therefore the cohomology
class of $H$ is independent of the splitting and it also characterises exact
Courant algebroids. The class of $H$ in $H^3(M, \mathbb{R})$ 
is called the Ševera class \cite{Severa-Weinstein} of $\mathcal{E}$.
In the split description of $E \cong TM \oplus T^*M$ the bracket is twisted
by the 3-form $H$ and takes the form

\be
\label{twisted dorfman}
[X + \xi, Y + \eta]_H = [X,Y] + {\cal L}_X \eta- \iota_Y d\xi + \iota_Y \iota_X H.
\ee

\bigskip

We shall now recall two propositions from
\cite{gcg-gualtieri}, that capture the idea of the
automorphism group of an exact Courant algebroid and how the B-field of emergent gravity satisfies the necessary condition in order to realize $e^b$ as a Courant algebroid automorphism in a splitting independent manner.

\smallskip

{\bf Proposition 2.1.1.} {\it Let $F\colon  E \rightarrow E$ be a vector bundle isomorphism covering the identity on $M$, that is orthogonal with respect to the inner product and preserves the anchor, i.e. for all $e_{1}, e_{2} \in \Gamma(E)$\\
1. $\left\langle e_{1}, e_{2}\right\rangle=\left\langle F e_{1}, F e_{2}\right\rangle$, \\
2. $\pi\left(e_{1}\right)=\pi \circ F\left(e_{1}\right)$,\\
then $F$ is a B-transform for some $b \in \Omega^{2}(M)$.}

\smallskip

On the other hand, such a B-transform does not necessarily preserve the Dorfman bracket. If $E \cong TM \oplus T^*M$ is equipped with the
H-twisted bracket we have

\be
\begin{aligned}
\left[e^{b}(X+\xi), e^{b}(Y+\eta)\right]_{H} &=\left[X+\xi+\iota_{X} b, Y+\eta+\iota_Y b\right]_{H} \\
&=[X+\xi, Y+\eta]_{H}+d\iota_{X} \iota_{Y} b-\iota_{Y} d \iota_{X} b+\iota_{X} d \iota_{Y} b \\
&=[X+\xi, Y+\eta]_{H}+\iota_{Y} \iota_{X} b+\iota_{[X, Y]} b \\
&=e^{b}[X+\xi, Y+\eta]_{H+d b}.
\end{aligned}
\ee

Therefore $e^b$ is an automorphism of the exact Courant algebroid $E$ if and
only if $b$ is closed. With this we are ready to describe $Aut(\mathcal{E})$.

\smallskip

{\bf Proposition 2.1.2.} {\it The automorphism group $Aut(\mathcal{E})$ of a Courant algebroid is a semidirect product fitting into the following short exact sequence
\be
\label{auto}
1 \rightarrow \Omega_{c l}^{2}(M) \rightarrow A u t(\mathcal{E}) \rightarrow \operatorname{Diff}_{[H]}(M) \rightarrow 1,
\ee
where $\operatorname{Diff }_{[H]}(M)$ is the subgroup of diffeomorphisms of $M$ preserving the Ševera class of $\mathcal{E}$ and $\Omega_{cl}^{2}(M)$ is the space of closed 2-forms on $M$.}

\smallskip 

In the case of emergent gravity, the $B$-field featured as an input is a closed 2-form, see
\cite{hsy-emergent, hsy-emergent2, lee1, lee2, lee3, hsy-quant}. Thus via the previous proposition we see that, in the context of emergent gravity, the B-field from which gravity emerges also defines a Courant algebroid automorphism. Later, after introducing generalized geometry, we will make precise the connection between the emergent metric and this Courant algebroid automorphism.

\smallskip

\subsection{Atiyah algebroid in Emergent Gravity}

We refer the reader to the page long diagrammatic exposition of emergent
gravity that was introduced by Yang \cite{hsy-emergent} by considering the 
deformation of a symplectic manifold $(M, b)$ where $b$ is a non-degenerate, closed 2-form on $M$.
We consider a line bundle $L$ over $(M, b)$ whose connection 1-form is denoted by $A = A_\mu (x) dx^\mu$ and the 
curvature $F$ of the line bundle is a closed 2-form, i.e., $dF=0$ and so locally can be expressed as $F = dA$. The line
bundle $L$ over $(M,b)$ admits a local gauge symmetry $\mathfrak{B}_L$ which acts on the connection $A$
as well as the symplectic structure $b$ as follows:
\be
\mathfrak{B}_L \colon  (b, A) \mapsto (b- d\Lambda, A + \Lambda),
\ee
where $\Lambda$ is an arbitrary 1-form on $M$.
The local gauge symmetry $\mathfrak{B}_L$ is known as the $\Lambda$-symmetry. 
This symmetry demands that the curvature $F = dA$ of $L$  appears only as the combination $\mathcal{F} \equiv b + F$ since the 2-form 
$\mathcal{F}$ is  gauge invariant 
under the $\Lambda$-symmetry. Since $d \mathcal{F} = 0$, this $\mathcal{F}$ can be considered as a ``dynamical" symplectic form for the symplectic manifold  $(M, \mathcal{F})$.
If $\det(1 + F \theta) \neq 0$ where $\theta \equiv b^{-1}$ is the
Poisson structure associated to $b$.
Therefore the electromagnetic field $F = dA$ manifests itself as the
deformation of a symplectic manifold $(M, b)$.
The local gauge symmetry $\mathfrak{B}_L$  induced by the symplectic structure on M, defines a bundle isomorphism 
$b\colon  TM \to T^* M$ by $X \mapsto %\Lambda =%
-\iota_X b$ where $X \in \mathfrak{X}(M)$ is
an arbitrary vector field such that the $\Lambda$ transformation  can be written as
\be
\mathfrak{B}_L \colon  (b, A) \mapsto \big(b+ \mathcal{L}_X b, A- \iota_X b \big),
\ee
 where $\mathcal{L}_X = d \iota_X + \iota_X d$ is the Lie derivative with respect to the vector field $X$.

\bigskip

 Emergent gravity naturally lives on the Atiyah algebroid of a torus fibration, where gauge fields and geometry are unified as algebroid data. Torus fibrations encode the topological and flux information necessary for nontrivial emergence, while the Atiyah algebroid provides the correct infinitesimal object replacing the tangent bundle. T-duality acts not on spacetime geometry alone but on the Courant algebroid built from the Atiyah algebroid, ensuring that emergent gravitational structures are transported consistently across dual backgrounds, even when the dual geometry becomes non-geometric.
The Atiyah algebroid of a torus bundle encodes the infinitesimal geometry of the fibration; it is an extension of the tangent bundle by a Lie algebra bundle, whose extension class records the topology of the torus bundle and provides the natural stage for T-duality and generalized geometry.

\bigskip

We begin by recalling some standard background on principal bundles and their associated Atiyah algebroids.
We will recall how in the presence of a connection on a principal bundle one can decompose the sections of the Atiyah algebroid in terms of Lie algebra valued forms. This will allow us to compute T-duals of generalized metrics in a coordinate free description.  These results are well known and can be found in standard references; see, for example, Kobayashi and Nomizu \cite{KN-Book} for a detailed treatment on principal bundles. In this work, we focus primarily on torus bundles, i.e. principal bundles with structure group $\mathbb{T}$.
Nevertheless, many of the definitions and constructions presented here extend naturally to the more general setting of principal bundles with compact, connected, semisimple, or abelian structure groups $G$. 
This means there is a non-degenerate
$\mathbb{T}$-invariant symmetric bilinear pairing on the Lie algebra 
$\mathfrak{t}$ of  $\mathbb{T}$. The Lie bracket on $\mathfrak{t}$  is defined as the Lie bracket of left invariant vector fields on $\mathbb{T}$ for sign convention.

\smallskip

Let $\pi\colon  M \rightarrow B$ be a principal $\mathbb{T}$-bundle. The fundamental vector fields on $M$ are defined as the vector field generated by differentiating the action of $\mathbb{T}$ on $M$. Fundamental vector fields span the {\it vertical} subbundle of $TM$ consisting of vectors that are in the kernel of $\pi_{*}.$ Notice that the action of $\mathbb{T}$ on $M$ lifts naturally to the tangent bundle $TM \rightarrow M$ by differentiation. With this in mind we define. 

\smallskip

{\bf Definition 2.2.1.} Let $\pi \colon  M \rightarrow B$ be a principal $\mathbb{T}$-bundle. Then the {\it Atiyah algebroid}
corresponting to $M$ is the vector bundle $\mathcal{A} = TM/\mathbb{T}$ over $B = M/\mathbb{T}$.

\smallskip

{\bf Proposition 2.2.2.} The Atiyah algebroid of $M$ is a Lie algebroid over $B$ with
surjective anchor induced by $\pi_{*} \colon  TM \rightarrow TB$. Moreover
$\mathcal{A}$ fits into the short exact sequence of Lie algebroids
\be
0 \longrightarrow \mathfrak{t}_M \stackrel{j}{\longrightarrow} \mathcal{A} \stackrel{\pi_*}{\longrightarrow} T B \longrightarrow 0,
\ee
where $\mathfrak{t}_M$ is the vector bundle over $B$ associated to the adjoint representation of $\mathbb{T}$ on $\mathfrak{t}$. The {\it sequence} is called the 
Atiyah sequence corresponding to the principal bundle $M$. For a proof of this proposition readers may consult \cite{GKP}.

\smallskip

 A differential form $\omega$ on $M$ is called invariant if the pullback by the right $\mathbb{T}$-action satisfies $R_{t}^{*} \omega=\omega$ for all $t \in \mathbb{T}$. The space of invariant differential
 forms is denoted by $\Omega^{\bullet}(M)/\mathbb{T}$. Sections of the Atiyah algebroid are identified with $\mathbb{T}$-invariant sections of
 $T M$ and hence the Atiyah sequence induces a filtration of $\Omega^{\bullet}(M)/\mathbb{T}$
\be
\label{filtration}
 \Omega^{\bullet}(B)=\mathcal{F}^{0} \subset \mathcal{F}^{1} \subset \ldots \subset \mathcal{F}^{n}=\Omega^{\bullet}(M)/\mathbb{T},
 \ee
 where $n$ is the dimension of $\mathfrak{t}$ and $\mathcal{F}^{i}=A n n\left(\wedge^{i+1} \mathfrak{t}\right)$ is the subspace of forms that vanish on $\wedge^{i+1} \mathfrak{t}$.

\smallskip

\subsection*{Connections}

A connection 1-form on the principal bundle $M$ is an {\it equivariant} Lie algebra valued 
1-form $A \in \Omega^{1}(M, \mathfrak{t})$, meaning that
$R_{t}^{*} A=A d\left(t^{-1}\right) A$,  for all $t \in \mathbb{T}$, where $A d$ is the adjoint action of $\mathbb{T}$ on the Lie algebra, such that
$\iota_{\psi(x)} A=X$ for all $x \in \mathfrak{t}$.
By equivariance, the connection 1-form descends to a right splitting $j:\mathfrak{t}_M\to \mathcal{A}$ of the
Atiyah sequence. Given a connection $A \in \Omega^{1}(M, \mathfrak{t})$ the Atiyah sequence
splits as $\mathcal{A} \cong {\mathfrak{t}}_{M} \oplus T B$. 
Then we can identify sections of $\mathcal{A}$ with $\mathbb{T}$-invariant sections of $T M$ which can be written as
\be
\begin{aligned}
T B \oplus {\mathfrak{t}}_{M} & \rightarrow T M \\
X+s & \mapsto X^{H}+j(s)=X^{H}+s,
\end{aligned}
\ee
where $X^{H}$ is the horizontal lift and $j$ is the map from the Atiyah sequence. We omit $j$ later, although one has to be careful with signs. 

A connection form $A$ on $M$ turns the filtration (\ref{filtration}) of
invariant differential forms into a decomposition. At degree $k$ it is given by
\be
\Omega^{k}(M)/\mathbb{T}=\bigoplus_{i=0}^{k} \Omega^{i}\left(B, \wedge^{k-i} \mathfrak{t}\right).
\ee
Therefore, if $a^{i}$ is some basis of $\mathfrak{t}$, any invariant differential $2$-form $b$  can be written as
\be
b=b_{2}+b_{1,i}\wedge a^{i}+b_{0,ij} \,
a^i\wedge a^j,
\ee
where $b_{2}, b_{1,i}, \,b_{0,ij}\, $ are
basic forms of degree 2, 1 and 0
respectively. Let us note that although in this section we focused on the differential forms, analogous results also holds for more general tensors.

\subsection{Reduction of the Courant Algebroid in Emergent Gravity}

In emergent gravity we start with an exact Courant algebroid $E$ defined over a smooth manifold $M$. We consider 
 a Lie group $\mathbb{T}$ acting on $M$ via diffeomorphisms. Assuming that the action is free and proper, $M$ acquires the structure of principal $\mathbb{T}$ bundle with compact connected structure group $\mathbb{T}$.
 The corresponding smooth quotient space is then given by $B \coloneqq  M/ \mathbb{T}$. We want to consider an action of $\mathbb{T}$ on the total space of $E \cong TM \oplus T^*M$ via automorphisms of $E$ and reduce 
it to a new non-exact Courant algebroid on the base $B$. Coming from the action of $\mathbb{T}$ we have Lie algebra homomorphism 
defining the infinitesimal action of $\mathfrak{t}$ on $\mathfrak{X} (M)$ via the adjoint action (Lie bracket) of vector fields.

\smallskip

{\bf Definition 2.3.1.} 
A lifted action of $\mathbb{T}$ on $E$ is a right action of $\mathbb{T}$ on $E$ via automorphisms covering the action of $\mathbb{T}$ on $M$. A lifted infinitesimal action of $\mathfrak{t}$ on $E$ is a Lie algebra homomorphism $\alpha\colon  \mathfrak{t} \rightarrow \operatorname{Der}(E)$ covering the infinitesimal action $\psi$ of $\mathfrak{t}$ on $T M$.

\smallskip

Given a lifted action of $\mathbb{T}$ on $E$ by differentiation we obtain a
lifted infinitesimal action of $\mathfrak{t}$ on $E$. Conversely if a
lifted infinitesimal action is obtained via differentiating a lifted action
of $\mathbb{T}$ we say that the lifted infinitesimal action integrates to an
action of $\mathbb{T}$ on $E$. First, we want to see whether under a group
action the invariant sections of an exact Courant algebroid could become a
Courant algebroid itself.

\smallskip

Let $\sigma\colon  M \rightarrow B$ be a principal $\mathbb{T}$-bundle and $(E,[\cdot, \cdot],\langle\cdot, \cdot\rangle, \pi)$ an exact Courant algebroid on $M$ so that $\mathbb{T}$ acts via diffeomorphisms preserving the Ševera class of $E$. As $\mathbb{T}$ is compact by averaging we may chose a splitting $E \cong T M \oplus T^{*} M$ such that the bracket is twisted by a $\mathbb{T}$-invariant 3-form $H \in \Omega^{3}(B)/\mathbb{T}$.

\smallskip

Then there is a natural extended action of $\mathbb{T}$ on $T M \oplus T^{*} M$ via
\be
\varphi(X+\xi)=\left(\begin{array}{cc}
\varphi_{*} & 0 \\
0 & \left(\varphi^{-1}\right)^{*}
\end{array}\right)\left(\begin{array}{c}
X \\
\xi
\end{array}\right)=\varphi_{*} X+\left(\varphi^{-1}\right)^{*} \xi.
\ee

\smallskip

By Proposition 2.1.2 and Equation (\ref{auto}), this action is an automorphism of the H-twisted Courant algebroid if and only if $\varphi^{*} H=H$ which is
 satisfied as we chose $H$ to be $\mathbb{T}$-invariant. Consequently if $M$ is a principal $\mathbb{T}$-bundle, the group action naturally lifts to an action on $E$  via Courant automorphisms.

\smallskip

Now we can define a new Courant algebroid over the base $B=M / \mathbb{T}$
of $M$. Firstly, $E / \mathbb{T}$ is a vector bundle over $B$ as the action
is free and proper on $M$ and acts via automorphisms on $E$. The sections of
$E / \mathbb{T}$ naturally identify with the $\mathbb{T}$-invariant
sections of $E$. These sections are closed under the Courant bracket and
their inner product is a $\mathbb{T}$-invariant function on $M$. Hence
$[\cdot, \cdot]$ and $\langle\cdot, \cdot\rangle$ descend to well defined
operations on $E / \mathbb{T} \rightarrow M / \mathbb{T} .$ Finally, to
define the anchor for our new Courant algebroid notice that the anchor $\pi$
of $E$ sends $\mathbb{T}$-invariant sections of $E$ to $\mathbb{T}$
-invariant sections of $T M$ which project to sections of $T(M / \mathbb{T})
.$ Therefore we can use $\pi$ as the anchor for $E / \mathbb{T}$
\be
\pi^{\mathbb{T}}\colon  E / \mathbb{T} \rightarrow T M / \mathbb{T} \rightarrow T(M / \mathbb{T})\cong TB.
\ee

\smallskip

It is easy to check that $\left(B, E / \mathbb{T}, \pi^{\mathbb{T}},[\cdot, \cdot],\langle\cdot, \cdot\rangle\right)$ satisfies the axioms of a Courant
algebroid over $B$. The reduced Courant algebroid is not exact, as the rank of $E / \mathbb{T}$ is too large but still transitive. Moreover, considering
the $\mathbb{T}$-invariant splitting $E \cong T M \oplus T^{*} M$ it is clear that $E / \mathbb{T}$ becomes a Lie algebroid over $B$ and 
$(T M / \mathbb{T})^{*} \cong T^{*} M / \mathbb{T} .$ Hence $E / \mathbb{T}$ fits into the short exact sequence
\be
0 \longrightarrow(T M / \mathbb{T})^{*} \stackrel{\frac{1}{2} \pi^{*}}{\longrightarrow} E / \mathbb{T} \stackrel{\pi}{\longrightarrow} T M / \mathbb{T} \longrightarrow 0.
\ee
Therefore, the resulting Courant algebroid is the double of the Lie
algebroid $T M / \mathbb{T}.$ 
%----
%It is easy to check that $\left(B, E / \mathbb{T}, \pi^{\mathbb{T}},[\cdot, \cdot],\langle\cdot, \cdot\rangle\right)$ satisfies the axioms of a Courant
%algebroid over $B$. The reduced Courant algebroid is not exact, as the rank of $E / \mathbb{T}$ is too large but still transitive. Moreover, considering
%the $\mathbb{T}$-invariant splitting $E \cong T M \oplus T^{*} M$ it is clear that $E / \mathbb{T}$ becomes a Lie algebroid over $B$ and since
%$(TM / \mathbb{T})^* \cong T^{*} B$. Hence $E / \mathbb{T}$ fits into the short exact sequence
%\be
%0 \longrightarrow(T M / \mathbb{T})^{*}\cong T^*B \stackrel{\frac{1}{2} \pi^{*}}{\longrightarrow} E / \mathbb{T} \stackrel{\pi^{\mathbb{T}}}{\longrightarrow} TM/\mathbb{T}\cong T B \longrightarrow 0
%\ee
%Therefore, the resulting Courant algebroid is the double of the Lie algebroid $T M / \mathbb{T} .$ This is what can be called simple reduction for
%emergent gravity.
%-----
This construction easily generalizes to any Courant algebroid $E$ over a principal $\mathbb{T}$-bundle $M \rightarrow B .$ Whenever the action of $\mathbb{T}$ lifts to $E$ via Courant algebroid automorphisms one can define a new Courant algebroid $E / \mathbb{T}$ over the base $B.$ We call this the simple reduction of $E$ in emergent gravity.

\section{Generalized Metric and Related Symmetries}
In this section we will introduce the generalization of Riemannian metrics on
exact Courant algebroids and see how they behave under reduction following
the notations of \cite{gcg-gualtieri}. The term generalized
metric was coined by Hitchin in \cite{2005hitchin}.

\smallskip

{\bf Definition 3.0.1} A generalized metric on $E$ is a self-adjoint, orthogonal bundle automorphism $\mathcal{G}\colon  E \rightarrow E$ which is positive definite in the sense that $\langle\mathcal{G} e, e\rangle>0$ for all non-zero sections $e \in \Gamma(E)$.
Orthogonality implies that $\mathcal{G} \mathcal{G}^{*}=\operatorname{Id}$, the identity automorphism, and together with the self-adjoint property we find that $\mathcal{G}^{2}=\operatorname{Id}$.
Therefore, a generalized metric also defines a decomposition of $E$ into its +1 and -1 eigenspaces which we denote by
\be
E_{-}=\operatorname{ker}(1+\mathcal{G}) \text { and } E_{+}=\operatorname{ker}(1-\mathcal{G}).
\ee
As $\mathcal{G}$ is positive definite the restriction of the inner product $\langle\cdot, \cdot\rangle$ to $E_{-}$ is negative definite and to $E_{+}$ is positive definite. Denote the projections to the subbundles $E_{\pm}$ as
\be
\Pi_{\pm}=\frac{1}{2}(1 \mp \mathcal{G})\colon  E \rightarrow E_{\pm}.
\ee
From this description it is clear that $E_{+}$ and $E_{-}$ are orthogonal with respect to the inner product.

\smallskip

Thus, if the inner product on a Courant algebroid has signature $(p, q)$
then a generalized metric is equivalent to a decomposition of $E$ into two
orthogonal subbundles $E=E_{+} \oplus E_{-}$ of rank $p$ and $q$
respectively, such that the restriction of the inner product to
$E_{+}\left(E_{-}\right)$ is positive (negative) definite. Clearly, defining
one of the bundles determines the other as $E_{+}=E_{-}^{\perp}$. One can
then recover the bundle morphism $\mathcal{G}$ by defining it as the
identity on $E_{+}$ and minus the identity on $E_{-}$.

\smallskip

For an exact Courant algebroid $(E,\langle\cdot, \cdot\rangle,[\cdot,
\cdot], \pi)$ on a manifold $M$ of dimension $n$ the inner product has
signature $(n, n)$. Therefore a generalized metric is equivalent to defining
a rank $n$ negative definite subbundle $E_{-} \subset E$ and we have the
decomposition $E=E_{+} \oplus E_{-}$ for $E_{+}=E_{-}^{\perp}$. Generalized
metrics can also be described from the point of view of the underlying
manifold utilising the bundle map $\lambda=\left(\pi_{-}\right)^{-1}\colon  T M \rightarrow E$. \footnote {
Given a generalized metric on a transitive Courant algebroid, the anchor
restricted to the subbundles $E_{-}$ and $E_{+}$
$\pi_{\pm}=\left.\pi\right|_{E_{\pm}}\colon  E_{\pm} \rightarrow T M $ is
surjective.} Thus we have the following proposition.

\smallskip

{\bf Proposition 3.0.2} A generalized metric $E_{-}$ on an exact Courant algebroid $E$ over the manifold $M$ is equivalent to a Riemannian metric $g$ on $M$ and an isotropic splitting $E \cong T M \oplus T^{*} M$ such that $E_{-}$ and $E_{+}$ are of the form
\bea
\begin{array}{l}
E_{-}=\{X-g X \mid X \in T M\}, \\
E_{+}=\{X+g X \mid X \in T M\}.
\end{array}
\eea

Note that while a usual metric is a reduction of the frame bundle from $GL(n)$ to the maximal compact subgroup $O(n)$, a generalized metric 
is a reduction from $O(n, n)$ to $O(n) \times O(n)$.

\smallskip

The symmetry group of the form (\ref{ip}) is the orthogonal group
\be
 O \left(T M \oplus T^{*} M\right)=\left\{A \in GL\left(T M \oplus T^{*} M\right) \mid\langle A \cdot, A \cdot\rangle=\langle\cdot,\cdot\rangle\right\}.
\ee
Since the bilinear form has signature $(n, n),$ we have $O\left(T M \oplus T^{*} M\right) \cong O(n, n)$. The Lie algebra
\be
O\left(T M \oplus T^{*} M\right)=\left\{\mathbb{Q} \in M \left(T M \oplus T^{*} M\right) \mid\langle \mathbb{Q}\cdot, \cdot\rangle+\langle\cdot, \mathbb{Q} \cdot\rangle=0\right\},
\ee
that consists of matrices of the form
\be
\mathbb{Q}=\left(\begin{array}{cc}
a & \theta \\
b &-a^{T}
\end{array}\right),
\ee
where
\bea
\begin{array}{ll}
a\colon  \mathfrak{X} (M) \rightarrow \mathfrak{X} (M), & a^{T}\colon  \Gamma\left(T^{*} M\right) \rightarrow \Gamma\left(T^{*} M\right), \\
\theta\colon  \Gamma\left(T^{*} M\right) \rightarrow \mathfrak{X} (M), & b\colon  \mathfrak{X} (M) \rightarrow \Gamma\left(T^{*} M\right),
\end{array}
\eea
satisfy $\theta^{T}=-\theta$ and $b^{T}=-b$. Hence we can think of $b$ as a 2-form $b \in \Gamma\left(\wedge^{2} T^{*} M\right)=\Omega^{2}(M)$ by $\iota_{X}b=b(X)$ and similarly $\theta$ as a bivector $\theta \in$ $\Gamma\left(\wedge^{2} T M\right)$. We thus see that generalized geometry, by its very nature incorporates 2-forms, i.e. B-fields. The finite transformations corresponding to $b$ and $\theta$ are given by
\bea
\begin{aligned}
e^{b} & \equiv\left(\begin{array}{cc}
1 & 0 \\
b & 1
\end{array}\right), & & e^{b}(X+\xi)=X+\xi+ \iota_X b, \\
e^{\theta} & \equiv\left(\begin{array}{cc}
1 & \theta \\
0 & 1
\end{array}\right), & & e^{\theta}(X+\xi)=X+\iota_{\xi} \theta+\xi.
\end{aligned}
\eea
We refer to $e^{b}$ as a B-transform and $e^\theta$ as a $\theta$-transform. The Courant bracket on $\Gamma\left(T M \oplus T^{*} M\right),$ which plays a similar role in generalized geometry as the Lie bracket on $\mathfrak{X} (M),$ is defined in \cite{Dirac_Courant}. 
Additional to B-transform and $\theta$-transform there exist diffeomorphism tranformations

\[ \mathcal{O}_N  \equiv\left(\begin{array}{cc}
N & 0 \\
0 & (N^T)^{-1}
\end{array}\right), \quad \mathcal{O}_N(X+\xi)= N(X)+ (N^T)^{-1}(\xi),\]
where $N:TM\to TM$ is a  diffeomorphism which is linear on the fibers, for example, a diffeomorphism induced from a diffeomorphism of the base $M$.

\smallskip

By the proposition stated above a generalized metric on $E$ is equivalent to the pair $(g, H)$ where $g$ is the Riemannian metric on $M$ and $H$ is the representative of the Ševera class \cite{Severa-Weinstein} of $E$ defined by the splitting.

2-forms act transitively on isotropic splittings of $E$ via the B-transform. Therefore, in an arbitrary splitting $E \cong T M \oplus T^{*} M$ the two subbundles defined by a generalized metric have the following form:
\bea
\label{graphs}
\begin{array}{l}
E_{+}=\{X+g(X)+b(X) \mid X \in T M\},\\
E_{-}=\{X-g(X)+b(X) \mid X \in T M\}.
\end{array}
\eea
Using this formulation, one can reconstruct $\tau\colon  \Gamma(E) \rightarrow \Gamma(E)$ as a $C^{\infty}(M)$-linear map of sections, such that $\tau^{2}=1 .$ For, elements $e_{1}, e_{2} \in \Gamma(E),$ we have
\be
\left(e_{1}, e_{2}\right)_{\tau}\coloneqq \left\langle\tau\left(e_{1}\right), e_{2}\right\rangle = \left\langle e_{1}, \tau\left(e_{2}\right)\right\rangle,
\ee
such that $(., .)_{\tau}$ is symmetric and defines a positive definite fiberwise metric, referred to as generalized metric on $E$. Also, since  $\tau^{2}=1,$ it is orthogonal and thus $\tau \in O(n, n)$. Moreover, we get two eigenbundles $E_{+}$ and $E_{-},$ corresponding to +1 and -1 eigenvalues of $\tau$.
Using $g$ and $b$ we can construct
\be
\tau(X+\xi)=\left(g-b g^{-1} b\right)(X)-g^{-1} b(X)+b g^{-1}(\xi)+g^{-1}(\xi),
\ee
for all $(X+\xi) \in \Gamma(E) .$ In the block matrix form,
\bea
\tau\left(\begin{array}{c}
X \\
\xi
\end{array}\right)=\left(\begin{array}{cc}
-g^{-1} b & g^{-1} \\
g-b g^{-1} b & b g^{-1}
\end{array}\right)\left(\begin{array}{c}
X \\
\xi
\end{array}\right).
\eea
The corresponding fiberwise metric $(\cdot, \cdot)_{\tau}$ can then be written in the block matrix form
\bea
\la{fiberprod}
(X+\xi, Y+\eta)_{\tau}=\left(\begin{array}{c}
X \\
\xi
\end{array}\right)^{T}\left(\begin{array}{cc}
g-b g^{-1} b & b g^{-1} \\
-g^{-1} b & g^{-1}
\end{array}\right)\left(\begin{array}{c}
Y \\
\eta
\end{array}\right).
\eea
The block matrix in formula (\ref{fiberprod}) can be written as a product of simpler matrices. Namely,
\bea
\left(\begin{array}{cc}
g-b g^{-1} b & b g^{-1} \\
-g^{-1} b & g^{-1}
\end{array}\right)=\left(\begin{array}{cc}
1 & b \\
0 & 1
\end{array}\right)\left(\begin{array}{cc}
g & 0 \\
0 & g^{-1}
\end{array}\right)\left(\begin{array}{cc}
1 & 0 \\
-b & 1
\end{array}\right).
\eea
Notice the generalized metric $\mathcal{G}$ (Equation \ref{fiberprod}) 
$$
\mathcal{G}=\left(\begin{array}{cc}
-g^{-1} b & g^{-1} \\
g-b g^{-1} b & b g^{-1}
\end{array}\right)
$$
is the B-transform of
\be
g \coloneqq \left(\begin{array}{ll}
g & 0 \\
0 & g^{-1}
\end{array}\right),
\ee
where we are abusing notation and using the same letter for the Riemannian metric and the diagonal generalized metric it defines, thus 
\be
\mathcal{G}= e^{b} g e^{-b}.
\ee
In what follows, we will continue adopting this abuse of notation. The distinction between these two realizations will be clear from the context, as in the discussion above. It is important to emphasize that the 2-form $b$ is not required to be closed, and this assumption will hold throughout the present work. For computational convenience, we assume that $b$ is globally defined, so that 
$H=d b$ is also globally defined and corresponds to a trivial class in integral cohomology, $[H]=0$. In the context of emergent gravity, we restrict attention to models with trivial H-flux. The case of nontrivial H-flux arises only after performing T-duality on the original theory, which will be the subject of the following section.

\section{A Plausible T-dual}

T-duality is a symmetry relating two a priori distinct string backgrounds, which nevertheless exhibit identical physical behavior. Its key implication is that type IIA and type IIB superstring theories are, in fact, two manifestations of a single underlying theory, distinguished by the interchange of winding and momentum modes, together with the replacement of the tori in spacetime by their duals in the target space. The local relationship between T-dual backgrounds, more precisely, the local transformation rules of the low-energy effective fields can be systematically described by the Buscher rules \cite{Buscher}. By contrast, the study of the global aspects of T-duality, such as the topological changes of the manifolds associated with T-dual backgrounds, was initiated in \cite{MR1265457} and has been developed extensively in \cite{TopTdual1, TopTdual2, TopTdual3}, where the \emph{algebraic and topological structures} underlying T-duality were systematically analyzed.

\bigskip

In the low-energy limit of type II string theory, the bosonic field content consists of a Riemannian metric $g$, a closed 3-form $H$, and a dilaton $\varphi$ all of which satisfy a set of modified Einstein equations. Remarkably, these equations possess a symmetry, T-duality,  that has no analogue in ordinary general relativity. 
This symmetry relates pairs of spaces $M$, and $\hat{M}$, which are principal torus bundles over a common base $B$, and can be characterized by the exchange of the Chern classes of the torus bundles together with topological data associated with the H-flux. 
It is now well established that T-duality admits a natural geometric formulation within the framework of generalized geometry \cite{guacav}. From this perspective, T-duality is realized as an isomorphism of twisted exact Courant algebroids. More precisely, given a principal $\mathbb{T}$-bundle $M\to B$
and the twisted Courant algebroid
$TM\oplus T^*M$ defined by the flux 
$H\in \Omega^3(M)$, we say that a T-dual background exists if there is a principal torus bundle $\hat{M}\to B$ together with a flux
$\hat{H}\in \Omega^3(\hat{M})$, 
such that $(TM\oplus T^*M,H) \cong (T\hat{M}\oplus T^*\hat{M},\hat{H})$.
Leaving aside the dilaton $\varphi$, the field content $(g,H)$ defines a 
generalized metric on the Courant algebroid $TM\oplus T^*M$.
Consequently, T-duality for the type II string equations can be understood as an isomorphism between generalized 
metrics. This geometric reformulation will serve as the foundation for our construction of a T-dual model of emergent gravity.

\subsection{Existence and Obstruction of T-dual Theories}

{\it A priori} the two T-dual bundles $M\to B$ and $\hat{M}\to B$ are completely unrelated, thus in order to compare them one needs to introduce the correspondence space 
\be
\la{correspondence}
M\times_B \hat{M}=\{(m,m')\in M\times M'\colon  \pi(m)=\pi'(m')\},
\ee 
i.e. the fiber product with respect to the canonical bundle projections. The two bundles fit into a diagram
$$
\begin{tikzcd}
& (M\times_B \hat{M}, p^*H-\hat{p}^*\hat{H}) \arrow[dl,"p"] \arrow[dr,"\hat{p}", swap] & \\
(M, H)   \arrow[dr,"\pi", swap] & & (\hat{M}, \hat{H})  \arrow[dl,"\hat{\pi}"] \\
& B &
\end{tikzcd}
$$

Notice that the correspondence space is a $\mathbb{T}$-bundle with base $M$ with action induced from the $\mathbb{T}$-action on $\hat{M}$, similarly it is a $\mathbb{T}$-bundle over $\hat{M}$ with the induced $\mathbb{T}$-action on $M$. Therefore there is an action of $\mathbb{T}\times \mathbb{T}$ on the correspondence space.

\bigskip

{\bf Definition 4.1.1}
Let $M$, $\hat{M}$ be two principal $\mathbb{T}$-bundles over a common base $B$ and let $H\in \Omega^3(M)$, $\hat{H}\in \Omega^3(\hat{M})$ be integral closed forms. We say $M$ and $\hat{M}$ are T-dual if 
$p^*H-\hat{p}^*\hat{H}=dK $,
for some $\mathbb{T}\times \mathbb{T}$-invariant form $K\in \Omega^2(M\times_B \hat{M})$ which is non degenerate as a form restricted to the subbundle $\mathfrak{t}_M\otimes \mathfrak{t}_{\hat{M}}$, where $\mathfrak{t}_M$ and $\mathfrak{t}_{\hat{M}}$ are the associated bundles to the adjoint action of $\mathbb{T}$ on  $\mathfrak{t}$ and $\hat{\mathfrak{t}}$.

\bigskip

%The non degeneracy condition on the form $F$ plays a crucial role in the description of T-duality as an isomorphism of Courant algebroids as we will see afterwards.

The obstruction to the existence of T-dual is well understood: let $M$ be a principal torus bundle, then $M$ is T-dualizable if and only if there exists a closed integral $\mathfrak{t}^*$-valued \footnote{$\mathfrak{t}^*$ denotes the dual of $\mathfrak{t}$, usually denoted by $\hat{\mathfrak{t}}$ which we avoid since we will be using the notation $\hat{\cdot}$ for T-dual structures later.} 2-form $\kappa \in \Omega^2(B,\mathfrak{t}^*)$ on $B$, such that the pair $(H,\kappa)$ satisfies
\begin{equation}
\label{eq: ObstructionTDuality}
    dH=0,\qquad \iota_X H=\pi^*\kappa(X),
\end{equation} 
for all $X \in \Gamma(\mathfrak{t}_M)$.
Some clarifying remarks are in order. First, in the context of T-duality, one is concerned only with phenomena defined up to (co)homology; accordingly, it suffices to consider relations up to (co)homology. Furthermore, since every 1-form is homotopic to a $\mathbb{T}$-invariant one, we may, without loss of generality, restrict our discussion to $\mathbb{T}$-invariant objects, as outlined in {Section 2.2}. Second, although 
T-duals are not strictly unique, there exists, once torsion is fixed, a unique 
T-dual  $\mathbb{T}$-bundle up to isomorphism. To find the torsion free T-dual to a given T-dualizable bundle $M$ one argues as follows: 
Since the $\mathfrak{t}^*$-valued 2-form $\kappa$ is integral it defines a class in $H^2(M,\mathbb{Z}^n)$ which, by the long exact sequence of cohomology induced by the exponential map, defines in turn a class
in $H^1(B,\mathbb{T})$, i.e. a $\mathbb{T}$-bundle $\hat{M}\xrightarrow{\hat{\pi}} B$. If we demand the bundle to be torsion
free, it follows that $\hat{M}$ is unique up to isomorphism. Consider connection
1-forms $a\in \Omega^1(M,\mathfrak{t})$ and
$\hat{a}\in \Omega^1(\hat{M},\mathfrak{t}^*)$ such that 
\be 
da= \pi^*\kappa,\quad d\hat{a}= \hat{\pi}^*\kappa,
\ee 
this always can be done, see for instance \cite{Kostant} for the case of circle bundles. Now define 
\be 
h\coloneqq  a\wedge d\hat{a}-H, 
\ee 
since $H, a$ and $\hat{a}$ are assumed to be invariant and $\iota_X 
h=d\hat{a}$, it follows that $h$ %is also invariant and $\iota_X h= \pi^*\hat{F}(H)-\iota_X H=0$ for all $X\in\mathfrak{t}$, thus $h$ 
is basic, i.e. the pullback of a form in $B$, abusing notation we will also
denote this form by $h\in \Omega^3(B)$. At last the T-dual NS flux can be
defined as 
\be
\hat{H}=da\wedge \hat{a}- h.
\ee
Once again we are slightly abusing notation by treating $da$ as basic form and hence regarding it as a form on base $B$. It is clear $H$ is a closed invariant form.
Moreover the couple
$(\hat{M},\hat{H})$ is a T-dual bundle to $(M,H)$, indeed
\be 
p^*H-\hat{p}^*\hat{H}^*=a\wedge d\hat{a}-da\wedge \hat{a}=-d(a\wedge \hat{a}),
\ee 
and restricted to the fibers $\mathfrak{t}_M\otimes \mathfrak{t}_{\hat{M}}$ the form $a\wedge  \hat{a}$ is just the canonical pairing between $\mathfrak{t}$ and $\mathfrak{t}^*$ therefore non-degenerate. In emergent gravity we focus on the situation where $H=db$ for a globally defined, closed B-field $b$. In this case the corresponding flux is trivial, $H=0$. We will return to this special case after first developing a formulation of T-duality within the framework of Courant algebroids and generalized geometry.

\subsection{T-duality as an Isomorphism of Courant Algebroids}

\smallskip

In the work by Calvancanti and Gualteri \cite{guacav}, T-duality is exposed in the
context of generalized geometry as an isomorphism of Courant algebroids associated to
the twisted generalized tangent bundles of $(M,H)$ and $(\hat{M},\hat{H})$
respectively. This description allows to transport generalized geometric data from one
T-dual bundle to another. 

\smallskip

We begin with the observation that upon choosing connections $a\in \Omega^1(M,\mathfrak{t})$  the bundles $(TM\oplus T^*M)/\mathbb{T}$ and $(T\hat{M}\oplus T^*\hat{M})/\mathbb{T}$ decompose as
\be 
(TM\oplus T^*M)/\mathbb{T}\cong TB\oplus \langle \partial \rangle \oplus T^*B \oplus \langle a \rangle, 
\ee 
where $\partial, \hat{\partial}$ are the vertical vector fields associated to the connection forms $a,\hat{a}$ respectively. In the case of a $n$-torus $\mathbb{T}^n$ the connection forms can be actually described by $n$ 1-forms $a=(a^1,\ldots a^n)$ and the vector fields $\partial=(\partial_1,\ldots \partial_n)$ are the dual vector fields satisfying $a^i(\partial_j)=\delta^i_{j}$. Analogously choosing a connection on the T-dual bundle $\hat{a}\in \Omega^1(\hat{M},\hat{\mathfrak{t}})$
yield a decomposition 
\be 
(T\hat{M}\oplus T^*\hat{M})/\mathbb{T}\cong TB\oplus \langle \hat{\partial }\rangle \oplus T^*B \oplus \langle \hat{a} \rangle. 
\ee 
The naive idea to define T-duality as a map of Courant
algebroids is as follows: one would like to pullback sections
of $(TM\oplus T^*M)/\mathbb{T}$ to sections of the Courant
algebroid of the correspondence space and then pullback along
the projection to $(T\hat{M}\oplus T^*\hat{M})/\mathbb{T}$ in
order to obtain a map 
$(TM\oplus T^*M)/\mathbb{T} \to (T\hat{M}\oplus T^*\hat{M})/\mathbb{T}$.
However, this naive procedure requires a careful formulation. On one hand, the pullback of vector fields is not well-defined; on the other hand, the pushforward of differential forms is meaningful only for basic forms. Consequently, to make this construction precise, one must exploit the additional structure available on the correspondence space.

In an intuitive picture, depicted in (Figure 2): 
start with a vector field  
$X\in \mathfrak{X}(M)/\mathbb{T}$ and a 1-form $\xi\in \Omega(T^*M)/\mathbb{T}$, 
one may lift the initial vector field to some vector field 
$\tilde{X}\in \Gamma(TM\times_B T\hat{M})$ which pushforwards to $X$
and pullback the form $\xi$ to $p^*\xi$; then do an adequate B-field
transformation, depending on $\tilde{X}$ and $K$, in order to change $p^*\xi$
to a basic form. At last push-forward both structures to get a section of
$(T\tilde{M}\oplus T^*\hat{M})/\mathbb{T}$ . It turns out that there is only one
$\tilde{X}$ such that the above procedure makes sense, and thus the map is well
defined. The concrete map is the following
\bea
\begin{aligned}
&\varphi\colon ( TM\oplus T^*M)/\mathbb{T}\to (T\hat{M}\oplus T^*\hat{M})/\mathbb{T},\\&\qquad  \varphi(X+\xi)=\hat{p}_*(\tilde{X})+p^*(\xi)-K(\tilde{X}), 
\end{aligned}
\eea
where $\tilde{X}\in \Gamma(T(M\times_B\hat{M}))$ is the unique lift of $X\in \mathfrak{X}(M)/\mathbb{T}$ satisfying  $\iota_Y \xi-K(\tilde{X},Y)=0$, for all
vector fields $Y\in \Gamma(\mathfrak{t}_M)$, recall these are the tangent fields to
the torus fiber on $M\to B$. Such a lift is well defined by non degeneracy
of the form $K$. It is worth mentioning that the previous condition is in
disguise the condition of $p^*\xi-K(\tilde{X})$ being basic, indeed it can be
rewritten as $\iota_Y(p^*\xi-K(\tilde{X}))=0$ and since both $\xi$ and $K$ are
$\mathbb{T}$-invariant then $\mathcal{L}_Y(p^*\xi-K(\tilde{X}))=0$, these are
exactly the condition for a form to be basic.    

\bigskip

\textbf{Theorem 4.2.1} (\cite{guacav})
For T-dual spaces $M$ and $\hat{M}$ the map $\varphi\colon  (TM\oplus T^*M)/\mathbb{T}\to (T\hat{M}\oplus T^*\hat{M})/\mathbb{T}$ defined above is an isomorphism of Courant algebroids. 

\smallskip 

 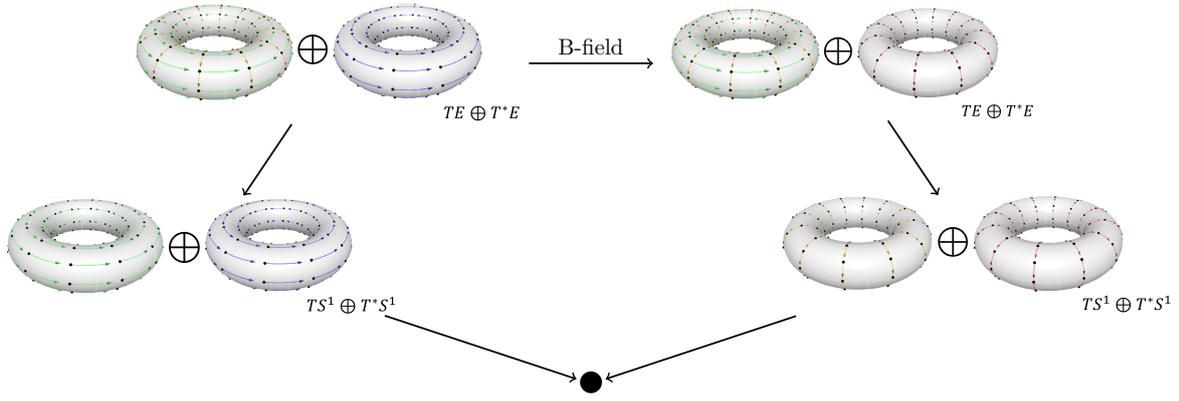
\begin{figure}[ht]
    \centering

    \begin{tikzpicture}
        % 1. Load the image
        \node[anchor=south west, inner sep=0] (image) at (0,0) {\includegraphics[width=0.9\textwidth]{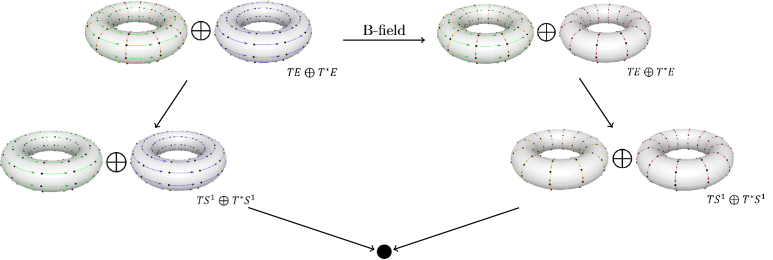}};

        \begin{scope}[x={(image.south east)},y={(image.north west)}]
            
            % 3. Add your math formulas
            \node[white] at (0.5, 0.7) {$E = mc^2$};  
            \node[fill=white] at (0.3, 0.22) {$TM\oplus T^*M$};
            \node[fill=white] at (0.98 ,0.22) {$T\hat{M}\oplus T^*\hat{M}$};
            \node[fill=white] at (0.485, 0.91) {B-tranform};
            \node[fill=white] at (0.4, 0.7) {$TE \oplus T^*E$};
            \node[fill=white] at (0.85, 0.7) {$TE \oplus T^*E$};
            \node[fill=white] at (0.54, 0.0) {$ B$};
            
        \end{scope}
    \end{tikzpicture}
 \caption{Schematic picture of T-duality. The base space $B$ is a point, the spaces of interest $M$ are circles and the correspondence space $E\coloneqq M\times_B \hat{M}$ is a torus. After lifting the vector field on $M$ one must do a B-transformation to make the 1-form vertical so it can be pushed forward.}
 
 \end{figure}
 
\bigskip 

\bigskip

\subsection{Trivial H-flux and Transport of Structures}
As previously noted, in our setting of interest, emergent gravity, H-flux 
is trivial, We therefore develop the Courant isomorphism explicitly in this case. When $H=0$ every $\mathbb{T}$-bundle is 
T-dualizable. Recall that the obstruction to the existence of T-dual torus bundles is encoded in a class $G\in\Omega^2(B,\mathfrak{t})$ satisfying (\ref{eq: ObstructionTDuality}).
In the trivial H-flux case it is straightforward to verify that
$G=0$ satisfies this condition. Consequently
the T-dual bundle exists, and by construction its Chern class vanishes ($G=0$), implying that the bundle is topologically trivial. The trivial bundle has a natural connection $\hat{a}$ given by
$\text{pr}_\mathbb{T}^*(a_{\mathbb{T}})$ of the unique invariant
connection\footnote{This is 
usually known as the Maurer Cartan connection and in the case of torus 
bundles is given by $(d\theta^1,\ldots,d\theta^n)$ where $d\theta$ is the unique normalized invariant form on $\mathbb{T}^1$, i.e. $d\theta_{g}(X)=\left(R_{g^{-1}}\right)_{*} X$.} $a_{\mathbb{T}}$ on $\mathbb{T}$  via the projection $\text{pr}_{\mathbb{T}}\colon B\times \mathbb{T}\to \mathbb{T}$. Since $a_{\mathbb{T}}$ is exact it follows $\hat{a}$ is also closed and by definition the T-dual flux is given by $\hat{H}=da\wedge \hat{a}$. We conclude:

\bigskip

\textbf{Proposition 4.3.1}
Let $M$ be a $\mathbb{T}$-bundle over $B$ with trivial H-flux, then the T-dual exists and is given by $B\times \mathbb{T}$ with $\hat{H}=da\wedge \hat{a}$. 

\bigskip

This proposition exhibits the usual slogan that T-duality exchanges between
trivial topology with non-trivial H-flux and non-trivial topology with trivial
H-flux. To find the Courant algebroid consider $\tilde{X}+\tilde{\xi}\in \Gamma(TM\oplus T^*M)/\mathbb{T}$,  a section and using the decomposition induced
by the connection $a=(a^1,\ldots, a^n)$, and the dual vector field
$\partial=(\partial_1,\ldots, \partial_n)$ decompose section as 
\be 
\tilde{X}+\tilde{\xi}=X+ f_ia^i+\xi+g^j\partial_j,\qquad X\in \Gamma(TB), \quad \xi\in \Gamma(T^*B),\quad f_i,g^j\in C^\infty(B).
\ee 
Now in this case $K= a\wedge \hat{a}=a_1\wedge \hat{a}_1+\cdots + a_n\wedge \hat{a}_n$ and the unique lift $\tilde{X}$ satisfying $\iota_{\partial_i}\xi-K(\tilde{X},\partial_i)=0$ $\text{for all }  i,$ (i.e. has no $a_i$ in its decomposition), is given by $\tilde{X}=X+ f^i \partial_i +g^i \hat{\partial_i}$, where $g^i=g_i$, that is we rise the index just to make sense of the summation convention. Upon doing the B-transformation

\bea 
\begin{aligned}
(\tilde{X}+p^*\tilde{\xi})-K(\tilde{X})&=(X+f^i\partial_i+g^i\hat{\partial_i}+\xi+g_i a^i)-  (g_i a^i-f_i\hat{a}^i) \\&=X+f^i\partial_i+g^i\hat{\partial}_i+\xi+f_i\hat{a}_i.
\end{aligned}
\eea 
At last, taking the pushforward along $\hat{p}$, namely fiber integration, the vector field in the $\partial_i$ variables we obtain $\varphi\colon (TM\oplus T^*M)/\mathbb{T}\to (T\hat{M}\oplus T^*\hat{M})/\mathbb{T}$ is given by
\begin{equation}
\label{eq:TdualEquation}
  X+ f_ia^i+\xi+g^j\partial_j\quad \mapsto\quad  X+ g_ia^i+\xi+f^j\partial_j.
\end{equation}
That is T-duality in this case is given by an interchange of the variables $f_i\leftrightarrow g^i$ which can be considered as an interchange of position and momenta coordinates.

\bigskip

Given the Courant algebroid isomorphism one may transport Dirac structures or generalized structures from the Courant algebroid $(TM\oplus T^*M)/\mathbb{T}$ to $(T\hat{M}\oplus T^*\hat{M})/\mathbb{T}$ in the following way: a Dirac (generalized complex) structure on $M$ is determined by $C\subset (TM\oplus T^*M)/\mathbb{T}$ a maximally isotropic subvariety of $(TM\oplus T^*M)/\mathbb{T}$ (the respective complexification), upon doing the Courant algebroid isomorphism the image maps again to a maximally isotropic subvariety of $(T\hat{M}\oplus T^*\hat{M})/\mathbb{T}$ with respect to the H-twisted bracket (the complexification) thus defining a H-twisted Dirac (generalized complex) structure for $\hat{M}$. 

\bigskip

To show how this procedure works in practice, and which will be of importance later, consider the generalized complex structure defined by a symplectic form $\omega$ on a $\mathbb{T}^2$-torus bundle $M$. Given the decomposition $T^*M=T^*B\oplus \langle a_1\rangle \oplus \langle a_2\rangle$ the symplectic structure decomposes as 
\be 
\omega=\omega_{0,12}\;a^1\wedge a^2+\omega_{1,1}\wedge a^1+\omega_{1,2}\wedge a^2+ \omega_2,
\ee 
where $\omega_{0,12}\in C^\infty(B)$, $\omega_{1,1},\omega_{1,2}\in \Omega^1(B)$ and $\omega_{2}\in \Omega^2(B)$. To ease notation we will write $\omega_0$ instead of $\omega_{0,12}$ in what follows. The generalized complex structure of $\omega$ is determined by its graph $C_\omega=\{X-i\,\iota_{X}\,\omega \,|\, X\in \mathfrak{X}(M)\}$, in which a general element is given explicitly in the decomposition on vertical and horizontal parts by
\begin{equation}
\begin{aligned}
\label{eq:Ap}
X&+f_1\partial_1\\&+f_2\partial_2\\&+\omega_2-f_1\omega_{1,1}-f_2\omega_{1,2}\\&+(-\omega_{0}f_2+\omega_{1,1}(X))a_1\\&+(\omega_{0}f_1+\omega_{1,2}(X))a_2
.\end{aligned}
\end{equation}

Under the Courant algebroid isomorphism a general element of $C_{\omega}$,  maps to an element 
\begin{equation}
\begin{aligned}
\label{eq:Ap201}
X&+\left[i\omega_0 f_2-i\omega_{1,1}(X)\right] \tilde{\partial}_{1}\\
&+ \left[-i\omega_0 f_1-i\omega_{1,2}(X)\right] \tilde{\partial}_{2}\\&+i\left[\iota_X(\omega_2)-\omega_{1,1}f_1-\omega_{1,2}f_2\right]\\&
+f_{1} \tilde{a}_{1}\\&
+f_{2} \tilde{a}_{2}.
\end{aligned}
\end{equation}
We wish to find a generalized complex structure of type 0, i.e. a symplectic form with a B-field (2-form) $\tilde{\omega},\tilde{b}$, such that the graph corresponding to $e^{b+i\omega}$ is given by the elements defined by $\varphi(C_\omega)$, i.e. elements of the form of Equation (\ref{eq:Ap201}). Again we consider a decomposition of $\tilde{\omega}$ and $\tilde{b}$ with respect to the decomposition $T^*M=T^*B\oplus \langle a_1\rangle+\langle a_2\rangle$. With such a decomposition a general element of the graph, of the unknown, $e^{\tilde{b}+i\tilde{\omega}}$ is:

\begin{equation}\begin{aligned}
\label{eq:Ap202}
X&+\tilde{f}_1 \tilde{\partial}_{1}\\
&+ \tilde{f}_2 \tilde{\partial}_{2}\\&-i\left[ \iota_X \tilde{\omega}_2-\tilde{\omega}_{1,1}\tilde{f}_1-\tilde{\omega}_{1,2}\tilde{f}_2+\iota_X(\tilde{b}_2)-\tilde{b}_{1,1}\tilde{f}_1-\tilde{b}_{1,2}\tilde{f}_2\right]\\&
+\left[i\tilde{\omega}_0 \tilde{f}_2-i\tilde{\omega}_{1,1}(X)-\tilde{b}_0\tilde{f}_2    +\tilde{b}_{1,1}(X) \right] \tilde{a}_{1}\\&
+\left[-i\tilde{\omega}_0 \tilde{f}_1-i\tilde{\omega}_{1,2}(X)+\tilde{b}_0\tilde{f}_1    +\tilde{b}_{1,2}(X)\right] \tilde{a}_{2}.\end{aligned}\end{equation}

Taking  $\tilde{f}_1=i\omega_0f_2-i\omega_{1,1}(X)$ and $\tilde{f}_2=-i\omega_0f_1-i\omega_{1,2}(X)$, the $\partial_1$ and $\partial_2$ components agree (notice $X+\tilde{f}_1\partial_1+\tilde{f}_2\partial_2$ is still a general element of $TM$). The condition $\tilde{C}_\omega$ to be the graph of $e^{\tilde{b}+i\tilde{\omega}}$  implies an equality of the expressions of (\ref{eq:Ap201}) and (\ref{eq:Ap202}), which in turn sets a system of equations that can be used to find the explicit description of $\tilde{\omega}$ and $\tilde{b}$.

Equation arising from the $\tilde{a}_1$ components and coefficients of $f_1$ is:
\begin{equation}
\label{eq:Ap203a}
1=\tilde{\omega}_0\omega_0+i\tilde{b_0}\omega_0 \quad \Rightarrow\quad \tilde{\omega}_0=\frac{1}{\omega_0},\quad \tilde{b}_0=0,
\end{equation}
where we used that $\tilde{\omega}$ is a symplectic form implying $\tilde{\omega}_0$ should be real. Analogously, the equations arising from the contraction with $X$ in the $a_1$ and $a_2$ components are:
\begin{equation}
\label{eq:Ap203b}
0=\tilde{\omega}_0\omega_
{1,2}(X)-i\tilde{\omega}_{1,1}(X)+\tilde{b}_{1,1}(X) \quad \Rightarrow\quad \tilde{\omega}_{1,1}=0,\quad \tilde{b}_{1,1}=\frac{-\omega_{1,2}}{\omega_0},
\end{equation}

\begin{equation}
\label{eq:Ap204}
0=-\tilde{\omega}_0\omega_
{1,1}(X)-i\tilde{\omega}_{1,2}(X)+\tilde{b}_{1,2}(X) \quad \Rightarrow\quad \tilde{\omega}_{1,2}=0,\quad \tilde{b}_{1,2}=\frac{\omega_{1,1}}{\omega_0}.
\end{equation}

At last using the result from (\ref{eq:Ap203b}) and (\ref{eq:Ap204}) to the
equation obtained from the contraction with $X$ on the basic 2-form we get:
\begin{equation}
\label{eq:Ap205a}
\tilde{\omega}_2(X)=\omega_2(X)-\frac{\omega_{1,1}(X)\omega_{1,2}}{\omega_0}+\frac{\omega_{1,2}(X)\omega_{1,1}}{\omega_0}-i\tilde{b}_2(X),
\end{equation}
which again using that $\tilde{\omega}_2$ should be real gives:
\be
\label{eq:Ap205b}
\tilde{\omega}_2=\omega_2-\frac{\omega_{1,1}\wedge \omega_{1,2}}{\omega_0},  \quad \tilde{b}_2=0. 
\ee

We conclude that the T-dual of a symplectic form $\omega$ on $M$ with symplectic
fibers ($\omega_{0}\neq 0$) is again a (twisted) symplectic form and is given by $\tilde{\omega}=\tilde{\omega}_0+\tilde{\omega}_{1,1}\wedge a^1+\tilde{\omega}_{1,2}\wedge a^2+\tilde{\omega}_2$ where each component is given by the equations  (\ref{eq:Ap203a})- (\ref{eq:Ap205b}). 
Previous works have already claimed that the T-dual of a symplectic structure is symplectic again, see e.g. \cite{guacav}. However, to the best of our knowledge, the explicit formulae for the T-dual symplectic form are new. Note that in the case that the symplectic form on the fiber vanishes at a point ($\omega_0|_p=0$), the T-dual symplectic form might not be defined. This phenomena could be attributed to a topology change, a mechanism similar to that established in \cite{MR1265457}.

\section{T-Duality and Emergent Gravity}
\subsection{Noncommutative U(1)-theory, Emergent Gravity and Generalized
Geometry}

Emergent gravity, as formulated in \cite{hsy-emergent, hsy-emergent2, lee1, lee2, lee3, cornalba}, arises as a consequence of the Darboux theorem in symplectic geometry. The Darboux theorem may be regarded as a symplectic analogue of the equivalence principle in Riemannian geometry. Intuitively, in the presence of an electromagnetic field strength $F$, the equivalence principle, when expressed through the Darboux theorem, ensures the existence of a local diffeomorphism that eliminates $F$. However, this procedure simultaneously introduces dynamical degrees of freedom into the metric, so that the spacetime metric effectively emerges from the combined data of a symplectic form $\omega$ and the electromagnetic field $F$. Within our framework for the T-dual formulation of emergent gravity, we consider a 
T-dualizable Dp-brane with constant background metric $g$ and constant background B-field. More precisely, we take $M$ to be a $\mathbb{T}$-bundle equipped with a connection $a=(a_1,\ldots,a_n)$ and trivial H-flux.
The manifold $M$ is assumed to be symplectic with a $\mathbb{T}$-invariant symplectic form $b$ determined by the background
B-field. In addition, we consider a line bundle with connection $(L\to M,A)$ such
that $dA=F$ (see the diagram provided in the glossary of the introduction).\\

We begin with a brief review of the relationship between non-commutative $U(1)$ gauge theory and the emergent gravity. 
In the semi-classical regime, a non-commutative gauge theory can be viewed as a deformation of an ordinary $U(1)$-theory by a Poisson structure. Concretely the data of such a theory
consists of a non-commutative line bundle $L^{NC}\to M$ equipped with a non-commutative
connection $A^{NC}$ and its corresponding field strength $F^{NC}$.
It has been already shown (see for example \cite{Witten, JurcoNonCom}), that this
description is equivalent to a
commutative $U(1)$ gauge theory on a line bundle $L\to M$ with a connection form $A$ (curvature $dA=F$) together with a 
Poisson bivector $\theta$. 
The relation between the commutative and non-commutative field strengths is encoded in the {\it Seiberg-Witten map} \cite{ncft-sw}, which provides a systematic way to express the non-commutative data in terms of the commutative fields and the Poisson structure:
\be
    \label{Seiberg-Witten}
    F^{NC}=(1+\theta F)^{-1} F. 
    \ee
The non-commutative action \cite{ncft-sw} is a Moyal deformation of the
Dp-brane with flat metric $g$ on a noncommutative target space with electromagnetic field strength $F^{NC}$
\be 
 \label{Seiberg-Witten1}
S^{NC}=\frac{1}{g_s^{2}} \int d^{n} x \sqrt{g} g^{i k} g^{j l} \operatorname{Tr} ((F^{NC})_{i j} * (F^{NC})_{k l}).
\ee
Here Tr denotes a trace over internal and not over spacetime coordinates. In a (noncommutative) gauge theory, the field strength $\left(F^{N C}\right)_{i j}=F_{i j}^a T^a$ is Lie algebra valued, where $T^a$ are the generators of the gauge group (e.g. $U(N)$, $SU(N)$). The star product $*$ acts on the spacetime dependence of the components, but the result is still a matrix in the gauge indices. Tr means the matrix trace over those gauge indices, ensuring the action be a gauge invariant scalar.
Schematically, $\operatorname{Tr}\left(F_{i j}^{N C} * F_{k l}^{N C}\right)=\left(F_{i j}^a * F_{k l}^b\right) \operatorname{Tr}\left(T^a T^b\right)$ with $\operatorname{Tr}\left(T^a T^b\right)=\frac{1}{2}\delta^{ab}$.\\

Upon using the dictionary of commutative and non-commutative structures dictated by the {\it SW map} \cite{ncft-sw}, and considering an
expansion in the Poisson structure as
$\theta=1+\hbar\theta^{(1)}+\mathcal{O}(\hbar^2)$,
the semi-classical non-commutative action (\ref{Seiberg-Witten1}) is  up to
first order equivalent to an action of a commutative Dp-brane with field
strength $F$ and emergent metric $G^\text{em}$:
\be 
S=\frac{1}{G_s^{2}} \int d^{n} x \sqrt{G^\text{em}} (G^\text{em})^{i k} (G^\text{em})^{j l} \operatorname{Tr} (F_{i j} F_{k l})+\mathcal{O}(\hbar^2).
\ee 
The emergent metric $G^\text{em}$ up to first order in the deformation parameter $\theta$
is given by
\be
    \label{Seiberg-Witten2}
    G^\text{em}=g(1+F\theta^{(1)}). 
\ee 

From the perspective of generalized geometry, the data of a commutative Dp-brane with background metric $g$  
and gauge field strength $F$ 
can be encoded in the generalized metric $g+F$.
The non-commutative deformation is then naturally captured by a
$\theta$-transformation where the deformation parameter is taken to be $\theta=b^{-1}$. 
With this identification, one can show that the effective open-string metric $G$ can be 
obtained uniquely from the combined data ($\theta$, $F$, $g$). Indeed if we start with a generalized metric $g+F$ upon doing a B-transform $e^{-F}$ we obtain another
generalized metric 
\be 
g \xrightarrow{e^{F}} g+F,
\ee 
which after performing a $\theta$-transform with $e^{-\theta}$ leads to a
generalized metric $G+\Phi$. This is in agreement with the open-closed
string duality. In this article we start from a closed string background
$(g, b)$, and pick a poisson bivector $\theta$ to determine the open string
variables $(G, \Phi)$:
\be 
g+F\xrightarrow{e^{-\theta}} G+\Phi, \quad \qquad G=g(1-F\theta)^{-1},\quad \qquad \Phi=0.
\ee 
In fact, upon doing a $\theta$-transformation by $(b+F)^{-1}$ to the generalized
metric described by $g+b$ we obtain a new generalized metric $\mathcal{G}^\theta$
described by the metric $G$ and the 2-form $F$  satisfying the equality
\bea 
    \label{thetaTransformequality}
    \begin{array}{c}
G=g-F g^{-1}F, \quad G \theta=-Fg^{-1},
\end{array}
\eea 
where we can find $G$ explicitly as
\be 
G=g+G\theta F \quad\Rightarrow\quad G=g(1-\theta F)^{-1},
\ee 
which up to first order expansion in $\theta$ is $g(1+\theta^{(1)} F)$ and is
exactly the emergent metric (\ref{Seiberg-Witten2}). 
In the framework of generalized geometry, emergent gravity can be understood as the result of performing a $\theta$-transformation, $e^{\theta}$, subsequent to a B-transformation, $e^{F}$. This perspective also provides a natural explanation for the Seiberg–Witten formula for the non-commutative field strength $F^{\mathrm{NC}}$. Specifically, after applying the transformation $e^{\theta}$, the Dirac structure defined by the 2-form $F$ is mapped to another Dirac structure that is again determined by a 2-form if and only if the matrix $(1+\theta F)$ is invertible. In such cases, the resulting 2-form is precisely $F^{NC} = (1+\theta F)^{-1}F$ recovering the familiar Seiberg–Witten relation.
As an illustrative example, we now turn to the case of flat compact spacetimes. These constitute the earliest and most extensively studied setting in which T-duality naturally emerges within the generalized geometric framework, and they provide a mathematically well-defined context for the operation.

\subsection{Flat Spacetimes}
The theory of emergent gravity was first developed in the context of trivial spacetimes \cite{lee1,lee2, lee3, Witten} and subsequently extended to more general geometries \cite{LRY, LRY2, LRY3}, where its topological interplay with the underlying spaces was also investigated. In the present work, we focus on emergent gravity theories in the setting of toroidal compactification, emphasizing their relation to T-duality. Specifically, in this subsection we analyze toroidally compactified D$p$-branes on backgrounds characterized by a flat metric $g$ and a constant B-field.\\

From the T-duality point of view developed in the previous sections the
framework is obtained by setting the base space $B$ as a point: in this case
the total space and its dual are diffeomorphic to a torus, 
$M\cong \mathbb{T}$ and $\hat{M}\cong \mathbb{T}$; the space of invariant
forms and multi-vector fields are given exactly by constant forms and
multi-vector fields on the space $M$ or
$\hat{M}$; there is no appearance of H-flux on the T-dual by Proposition
{4.3.1} and the fact that $da=0$, since the exterior derivative of any form on the point manifold
vanishes. Moreover we assume the B-field is closed and invertible which
implies $M$ is a symplectic manifold with symplectic form $B$, thus an even
dimensional torus. 
We can describe the transport of generalized metrics on $M$ concretely via the Courant algebroid isomorphism. Let $\mathcal{G}$ be a generalized metric on $M$ described by the pair $(g,b)$
(as in Section 3). The isotropic subbundle defining this generalized metric is the eigenbundle
$E_{-}=\{X-g(X)+b(X) \mid X \in T M\}$. 
Using the decomposition of
$TM\oplus T^*M$ induced by the connection, 
elements of the eigenbundle take the form:
\be
X^k\partial_k +[(g+b)_{ij} a^i\otimes a^j](X^k\partial_k)= X^i \partial_i +(g+b)_{ik}X^k a^i. 
\ee 
Under the Courant algebroid isomorphism, such elements are mapped to  
\be 
X_i a^i + (g+b)^i_{k}X^k \partial_i= \tilde{X}^i \partial_i +{(g+b)^{k}_i}^{-1}\tilde{X}_k a^i.  
\ee 
Here we have defined $\tilde{X}^i=(g+b)^i_{k}X^k$ with the dummy indices used freely for raising and lowering. This reflects the emergent T-duality exchange 
$a^i\leftrightarrow \partial_i$.
We conclude that under the Courant isomorphism the generalized metric is
transformed as $\varphi\colon \mathcal{G}\mapsto \mathcal{G}^{-1}$, where we consider
this inversion as the block matrix inversion plus inversion of $g$ and $b$. This
T-duality action on generalized metrics has already been considered in the
literature, see for instance \cite{Porrati} for an extensive survey. The
concrete relationship between the metric $g+(b+F)$ and the T-dual counterpart
$\hat{g}+\widehat{b+F}$ is given by comparing the matrix components of the
corresponding generalized metrics 
\be 
\label{eq:Tdual}
    \hat{g}=g-(b+F)g^{-1}(b+F),\qquad \hat{g}\widehat{(b+F)}^{-1}=-(b+F)g^{-1} .
\ee 
We have shown why it is natural to consider the appearance of Poisson structures
as the counterpart of B-tranformation upon T-dualizing.\\

Consider the decomposition of $\mathcal{G}=g+b+F$ as 
$e^{b+F}g$, now T-duality acts on the map $e^{b+F}$ by conjugation, 
$\varphi \cdot e^{B+F} \cdot \varphi^{-1}$ which is a $\theta$-transform given by
$e^{(b+F)^{-1}}$. 
The $\theta$-transformation by $(b+F)^{-1}$ acts on the previously defined generalized metric $g$, yielding a new generalized metric $\mathcal{G}^{\theta}$. This transformed structure is characterized by the metric $G$, a vanishing 2-form $\Phi = 0$, and a Poisson bivector $\theta$, which together satisfy the following relations:
\bea 
    \begin{array}{c}
G=g-(b+F) g^{-1} (b+F), \qquad G \theta=-(b+F) g^{-1}.
\end{array}
\eea

Observing that the T-dual of the generalized metric $g$ is again $g$, and comparing this result with Equation (\ref{eq:Tdual}), we deduce the following equality:
\be 
e^{\theta}\varphi(g)=\varphi\left(  e^{b+F} g\right).
\ee

Thus {\it T-duality acts on B-transform by remolding them into
$\theta$-transformation}. 
Within the framework of generalized geometry, T-duality provides a natural and unified perspective on the well-known correspondence between open strings in the presence of a nonzero B-field and closed strings described by non-commutative gauge theories with the deformation parameter given by $\theta = (b+F)^{-1}$ \cite{Witten, Chu, Chu2, Hull}.\\
 
 \smallskip

%Indeed on the matrix representation of generalized metric 
%$$ \begin{aligned}\begin{pmatrix}
%1 & 0 \\-\theta & 1 
%\end{pmatrix}\begin{pmatrix}
%g & 0 \\ 0 & g^{-1} 
%\end{pmatrix}\begin{pmatrix}
%1 & \theta \\ 0 & 1 
%\end{pmatrix}&=\begin{pmatrix}
%1 & 0 \\-\theta & 1 
%\end{pmatrix}\begin{pmatrix}
%1 & B \\ 1 & 0 
%\end{pmatrix}\begin{pmatrix}
%g & 0 \\ 0 & g^{-1} 
%\end{pmatrix}\begin{pmatrix}
%1 & 0 \\-B & 1 
%\end{pmatrix}\begin{pmatrix}
%1 & \theta \\ 0 & 1 
%\end{pmatrix}\\&=\begin{pmatrix}
%1 &-B \\-\theta & 0 
%\end{pmatrix} \begin{pmatrix}
%g & 0 \\ 0 & g^{-1} 
%\end{pmatrix}\begin{pmatrix}
%1 &-\theta \\-B & 0 
%\end{pmatrix}\\&=\begin{pmatrix}
%0 & 0 \\ 0 &  0
%\end{pmatrix}\end{aligned}$$
We have thus far examined how non-commutative structures on strings and branes can be understood through the lens of T-duality within the framework of generalized geometry. We have already reviewed emergent gravity within the generalized geometry framework.  Our goal now is to understand the T-dual counterpart of emergent gravity using the generalized geometry. Specifically, we aim to characterize the T-dual of an emergent metric and establish whether it remains an emergent metric. As reviewed in the previous section, emergent gravity, in generalized geometry, is described by the composition of a $\theta$-transformation and a B-transformation, namely $e^{\theta} \cdot e^F$,
where $\theta = b^{-1}$ denotes the inverse of the background symplectic B-field. Upon performing T-duality, the emergent gravitational description transforms accordingly. Namely, as described above, T-duality turn the B-field transformation into a $\theta$-transformation, while the $\theta$-transformation becomes a B-field transformation. More concretely, the T-dual relationship can be expressed in the form of the following commutative diagram at the level of generalized metrics:
$$\begin{tikzcd}
g \ar[d, "e^F"] \ar[r,"\varphi"] & g\ar[d,"e^{F^{-1}}"]\\
g+F \ar[d,"e^\theta"]\ar[r, "\varphi"] & (g+F)^{-1}\ar[d,"e^{\theta^{-1}}"] \\
G \ar[r,"\varphi"]\ar[r, "\varphi"] & \hat{G} 
\end{tikzcd}. $$
Notice that the right side of the diagram, corresponding to the T-dual, does not describe an emergent gravity since it is given by a B-transform
after a $\theta$-transform, while emergent gravity is described by such
transformations in the inverse order. To get around this problem we use the fact that any symmetry of the generalized metric can be written in a standard form $e^{F^{\prime}} \mathcal{O}_{N} e^{\theta^{\prime}}$ see \cite{czech2}, thus
the composition $e^{\theta} \cdot e^{F}$ can be rewritten as
\be
\label{LHS-diagram}
e^{\theta}\cdot e^{F}=e^{F^{\prime}} \mathcal{O}_{N} e^{\theta^{\prime}},
\ee
for some $F',\theta'$ and $N$. One can compute that these are explicitly given by 
\be
\theta^{\prime}=(1+\theta F)^{-1} \theta=\theta(1+F \theta)^{-1},\\
\ee 
\be
F^{\prime}=F(1+\theta F)^{-1}=(1+F \theta)^{-1} F, \\
\ee
\be
N=1+F\theta.\\
\ee 
where the map $\mathcal{O}_N$ of Courant algebroids is a map induced by a diffeomorphism
$f_N$ of the manifold $M$ (which in this case is diffeomorphic to
$\hat{M}$) which on the Poisson structures induces an isomorphism
$(M,\theta)\cong (M',\theta')$. This diffeomorphism is obtained via the Moser trick and can be considered as a Darboux 
coordinate transformation which is the crux of emergent gravity, 
it is in fact the diffeomorphism mapping between the symplectic structures $b=\theta^{-1}$ and $b'=\theta'^{-1}$ which is at
the heart of our construction. 
For a detailed exposition of this result we refer the readers to \cite{czech}. After T-dualizing (\ref{LHS-diagram}) we obtain the equality
\bea 
\begin{aligned} \varphi e^\theta \cdot e^F \varphi^{-1}&=\varphi e^{F'} \mathcal{O}_N e^{\theta'} \varphi^{-1}\\&= \varphi e^{F'} \varphi^{-1}\varphi \mathcal{O}_N \varphi^{-1}\varphi  e^{\theta'}\varphi^{-1}\\&= e^{(F')^{-1}}\mathcal{O}_N e^{(\theta')^{-1}}.
\end{aligned}
\eea 
Therefore, up to a change of coordinates, given by the diffeomorphism
$f_N$, we conclude that the T-dual picture to emergent gravity is given
by the background electromagnetic field $\hat{F}$ and a
non-commutative structure with deformation parameter $\hat{\theta}$ given by 
\bea 
\label{hatrelation}
\begin{aligned} &\hat{F}\coloneqq (\theta')^{-1},\\&\hat{\theta}\coloneqq (F')^{-1}.
\end{aligned}
\eea 
Furthermore, if there exists a complex line bundle $\hat{L} \to \hat{M}$ equipped with a connection $\hat{A}$ such that $\hat{F} = d\hat{A}$ then, in this situation, one may regard $\hat{F}$ as the field strength of a $U(1)$ gauge theory. In this context, it is naturally interpreted as the electromagnetic field strength of the `T-dual theory.'\footnote{In mathematics, the existence of such an $\hat{A}$ is known as 'prequantizability'. This condition depends on certain cohomological obstructions of $\hat{F}$, namely it is required that $\hat{F}$ defines an integral cohomology class $[\hat{F}]\in H^2(M,\mathbb{Z})$. Since there exist a diffemorphism $f_N$ mapping $b'\to b$, or equivalently, $\theta'\to\theta$, the condition is equivalent to $[b]\in H^2(M,\mathbb{Z})$, i.e. that the original background B-field could be interpreted as the field strength of a $U(1)$-theory.} Notably, in this dual description, the T-dual electromagnetic field is associated with the original B-field $b$, whereas the T-dual non-commutativity parameter is determined by the original electromagnetic field strength $F$. We thus conclude that, in a flat spacetime with the T-dual description of emergent gravity remains an emergent gravity theory. However, as we shall explore in the following section, this equivalence may fail in curved backgrounds, where the base manifold is nontrivial and the generalized geometric structure becomes more intricate.\footnote{In \cite{RS} we have provided examples of T-duality of emergent gravity: first on self T-dual tori; and later on more complicated self T-dual nilmanifolds. In all those examples, the T-dual torus bundles have zero H-flux on both sides.} 

\subsection{Non-flat Spacetimes}
In the presence of a nontrivial base manifold, the 
T-dual geometry generally acquires a nonzero H-flux 
whose structure is determined by the curvature of the original torus bundle connection. This background flux modifies the underlying Courant algebroid: the Dorfman bracket is twisted by the H-flux and maximally isotropic subbundles correspond not to ordinary Dirac or generalized complex structures, but to their H-twisted counterparts. 
Consequently, the T-dual generalized metric is naturally twisted by 3 form H-flux. In contrast to the flat case, the presence of this twist obstructs a direct interpretation of the 
T-dual theory within the emergent gravity framework - one cannot simply map an emergent gravity description to another emergent gravity description under T-duality.\\

There are two possible strategies to circumvent this difficulty. One may restrict attention to $\mathbb{T}$-bundles with flat connection, ensuring that the T-dual background remains free of 
H-flux. However, this severely limits the class of geometries under consideration, and even in this restricted setting the construction of a Courant algebroid isomorphism for 
$dB\neq 0$ is highly nontrivial, rendering explicit computations cumbersome. Alternatively, one may adopt a fully covariant perspective by treating the original and 
the T-dual torus bundles on equal footing in the presence of H-flux. This approach remains an active area of investigation, with significant recent progress in both the physics and mathematics literature toward a unified understanding of 
T-duality and emergent gravity in flux backgrounds.

\smallskip

A general formula was derived for the topology and H-flux  of the T-dual for
a type II compactification in \cite{TopTdual1, TopTdual2, TopTdual3}. Summarizing, topological
T-duality uses the following topological data: A principal $U(1)$-bundle
$\pi\colon  M \rightarrow B,$ together with a pair of cohomology classes $(F, H)$
The class $F \in H^{2}(B, \mathbb{Z})$ is the first Chern class, and
determines the isomorphism class of the bundle, whereas the class 
$H \in H^{3}(M, \mathbb{Z})$ is the cohomology class of the curvature of the
B-field. T-duality intermixes $F$ and $H$. For an account of 
this topological version of T-duality between pairs of principal
$U(1)$-bundle (also extended for principal $\mathbb{T}^n$-bundles) equipped with
degree-3 integral cohomology class from a twisted K-theory point of view 
one may look at the works of Bunke et. al \cite{bunke,bunke2}.

\smallskip
An alternative mathematical perspective on the action of 
T-duality on generalized complex structures, both for vector spaces and torus bundles with trivial 
H-flux, has been developed by Ben-Bassat \cite{Ben}.
Subsequently,
Gualtieri and Cavalcanti \cite{GC} extended this analysis to the fully general setting of T-duality with nontrivial H-flux for higher-rank affine torus bundles, under the condition that
$\iota_X \iota_Y H = 0$ $\text{ for all }  X, Y$ tangent to the fibers.
It is noteworthy that in the earlier works of Mathai and Rosenberg \cite{MR, MR2,MR3}, this condition was not imposed; instead, they formulated 
T-duality by interpreting the dual space as a non-commutative geometry.
In particular, they demonstrated that T-dualizing along a $\mathbb{T}^2$
with non-vanishing H-flux produces a fibration by non-commutative tori.
Building on this Grange and Nameki \cite{Grange-Nameki} identified such non-commutative torus fibrations with the open string description of T-folds, and further embedded the non-geometric T-dual of a $\mathbb{T}^3$ with constant
H-flux into a generalized complex $\mathbb{T}^6$.

\smallskip

For further discussion of nontrivial H-flux and its relation to noncommutative gerbes, see \cite{ aschieri, aschieri2}. The situation becomes even subtler in the heterotic string, where T-duality differs from the conventional case: the dual background necessarily carries a gauge bundle with connection, leading to modified topological constraints for the existence of a T-dual. In this setting, the 
H-flux is no longer closed and thus fails to define a globally well-defined gerbe. A systematic treatment of these issues is presented in \cite{Baraglia-Hekmati}.

\smallskip

The geometry of double field theory \cite{vaisman-double} exhibits a structural similarity to that of emergent gravity on Calabi–Yau manifolds. In particular, Yang has recently shown that emergent gravity provides a natural framework for understanding mirror symmetry, where the doubling of six-dimensional manifolds arises as a consequence of Hodge theory applied to the simultaneous deformation of symplectic and dual symplectic structures in six dimensions \cite{hsy-mirror}. While these connections offer deep insights into the interplay between generalized geometry, mirror symmetry, and emergent gravity, a detailed exploration of these ideas lies beyond the scope of the present work and will be pursued in future investigations. 

\smallskip

Nevertheless, it remains possible to obtain an explicit expression for the generalized metric by following the procedure outlined in Section 4.3, even though an emergent gravity interpretation of the 
T-dual theory may no longer be available. To demonstrate this example in a concrete case we will consider
a spacetime $M$ which is a $\mathbb{T}^2$-fibration over an arbitrary base
manifold $B$. To find the T-dual generalized metric we first decompose,
according to the direct sum
$T^*M/\mathbb{T}\cong T^*B\oplus \langle a_1\rangle\oplus \langle a_2\rangle $ and 
$TM/\mathbb{T}\cong  TB\oplus \langle \partial_1\rangle\oplus \langle \partial_2\rangle $
given by the connection $a=(a_1,a_2)$, the background flat metric $g$, the
electromagnetic field 2-form $F$ and the Poisson structure $\theta$
describing the emergent gravity. In the $\mathbb{T}^2$-case these
decomposition are given by
\bea 
\label{decomposition}
\begin{aligned}
g&=a_1\odot a_1 +a_2\odot  a_2+g_2, \\
F&=F_{0,12}\;a_1\wedge a_2 + F_{1,1}\wedge a_1+F_{1,2}\wedge a_2+F_2,\\
\theta&=\theta_{0,12} \;\partial_1\wedge \partial_2 + \theta_{1,1}\wedge \partial_1+\theta_{1,2}\wedge \partial_2+\theta_2.
\end{aligned}
\eea 

In this decomposition we used $\odot$ to denote the symmetric tensor products and we have used 2 under-scripts for notational purpose, where the first under-script denotes the degree of the
form on $B$ and the second the connection which it is paired to, for example
$F_{1,2}$ is a basic 1-form and may be considered as $F_{1,2} \in
\Omega^1(B)$ and in the decomposition of $F$ appears wedged to $a_2$. Notice that $F_2$ and $g_2$ are 2-tensors in the base which defines a B-field and a Riemannian metric on $B$ respectively.  Notice also that the special
decomposition of $g$ has been used due to the fact that it is flat. Given
the previous decomposition and recalling $\partial_i(a^j)=\delta_i^j$ one
finds decomposition of $F\theta$ (considered as a (1,1)-tensor) to be 
\begin{equation}
\label{eq:EmergentMetric1}
\begin{split}
F\theta=&(F_{0,12}\;\theta_{0,12}+\theta_{1,1}(F_{1,1}))a_1\otimes \partial_1+(F_{0,12}\;\theta_{0,12}+F_{1,2}(\theta_{1,2}))a_2\otimes \partial_2\\
&+F_{1,1}(\theta_{1,2})a_1\otimes \partial_2 +F_{1,2}(\theta_{1,1}) a_2\otimes \partial_1\\
&+a_1\otimes[F_{0,12}\;\theta_{1,2}+\theta_2(F_{1,1})]+a_2\otimes [-F_{0,12}\;\theta_{1,1}+\theta_2(F_{1,2})]\\&+[\theta_{0,12}\;F_{1,2}+F_2(\theta_{1,1})]\otimes \partial_1+[-\theta_{0,12}\;F_{1,1}+F_2(\theta_{1,2})]\otimes \partial_2\\&+F_{1,1}\otimes \theta_{1,1} +F_{1,2}\otimes \theta_{1,2}+F_2\theta_2,
\end{split}
\end{equation}
where for example $\theta_2(F_{1,2})$ means the contraction of the bivector
field $\theta_2$ with the basic 1-form part of $F$. Using the
decomposition in (\ref{eq:EmergentMetric1}) the emergent metric $G$ (given
by Equation (\ref{Seiberg-Witten2})) is described by the decomposition
\be 
G=G_{0,11}\;a_1\odot a_1+G_{0,22}\;a_2\odot  a_2+G_{0,12}\;a_1\odot  a_2+G_{1,1}\odot  a_1+G_{1,2}\odot  a_2+G_2,
\ee 
where $\odot$ denotes the symmetrized tensor product, and the components are given by

\begin{equation}
\begin{split}
\label{eq:EmergentMetric2}
&G_{0,11}=1+F_{0,12}\;\theta_{0,12}+\theta_{1,1}(F_{1,1}),\\& G_{0,22}=1+F_{0,12}\;\theta_{0,12}+\theta_{1,2}(F_{1,2}), \\
    &G_{0,12}=F_{1,1}(\theta_{1,2}),\\
    & G_{1,1}=\theta_2(F_{1,1})+F_{0,12}\;\theta_{1,2},\\
    & G_{1,2}=\theta_2(F_{1,2})-F_{0,12}\;\theta_{1,1},\\
    & G_2=1+F_{1,1}\odot\theta_{1,1}+F_{1,2}\odot \theta_{1,2}+F_2\theta_2,
\end{split}    
\end{equation}
where we assumed $\theta F=F\theta$ to ensure that $G$ is symmetric.

One may now perform the 
T-duality transformation of the emergent metric by employing the Courant algebroid isomorphism associated with 
T-duality. The detailed derivation of the 
T-dual of an arbritary generalized metric on a 
$\mathbb{T}^2$-fibration
is presented in the appendix, where we derive a generalization of the classical Buscher rules, which concerns emergent gravity on a $\mathbb{T}^1$-bundle, to the $\mathbb{T}^2$-fiber case. Here, for convenience, we simply state the final results for the case of an emergent metric and a trivial $B$-field.

\begin{equation}
\label{eq:EmergentMetric2dual}
 \begin{aligned}
&\widetilde{G}_{0,11}=\frac{G_{0,22}}{G_{0,11}G_{0,22}-G_{0,12}^2},\quad && \widetilde{G}_{0,22}=\frac{G_{0,11}}{G_{0,11}G_{0,22}-G_{0,12}^2},\qquad &\widetilde{G}_{0,12}=\frac{G_{0,12}}{G_{0,11}G_{0,22}-G_{0,12}^2},\\
&\widetilde{G}_{1,1}=0,&& \widetilde{G}_{1,2}=0,\\
&\mathrlap{\tilde{G}_2=G_2+\frac{G_{0,12}(G_{1,1}\odot G_{1,2})-G_{0,11}( G_{1,2}\odot G_{1,2})-G_{0,22}( G_{1,1}\odot G_{1,1})}{G_{0,11}G_{0,22}-G_{0,12}^2}.}
 \end{aligned}
\end{equation}
 
 %The formulas in Equation \ref{eq:EmergentMetric2dual} can be considered as the result of applying the generalized $\mathbb{T}^2$-fiber Buscher rules to the  case of an emergent metric. 
 Let us mention that in this case, the T-dual of an emergent metric does not necessarily constitute an emergent gravity theory. This is demonstrated by explicit calculations in \cite{RS}, where the authors examine an emergent metric on $\mathbb{T}^2 \times N$ (with $N$ being a 4-dimensional nilmanifold). In this specific instance, the resulting T-dual metric fails to satisfy the criteria for an emergent metric \cite[Section 4.2]{RS}.
 Thus, for a non-flat spacetime, the above equations may be regarded as providing the defining framework for 
T-dual emergent gravity on a 
$\mathbb{T}^2$-fibration.

\smallskip

\section{Conclusion}
To conclude, let us summarize the key achievements of this work. We have developed a fully generalized geometric framework for emergent gravity, building it systematically from first principles, and carefully described the reduction of the exact Courant algebroid in this setting. With this foundation in place, we turned to the search for a T-dual formulation of the theory. Employing the Gualtieri–Cavalcanti (GC) isomorphism between Courant algebroids, we demonstrated how geometric structures can be consistently transported between torus bundles, revealing the dual picture in action.

\smallskip 

Most of Sections 2 and 3 serve a pedagogical purpose, offering a concise review of the relevant background, with the exception of Subsections 2.2 and 2.3, where we present new insights on the role of the Atiyah algebroid in emergent gravity and perform the reduction of the exact Courant algebroid in this context. The subsequent sections are devoted to our main results. In Section 4, we first revisit the Bouwknegt–Evslin–Hannabuss–Mathai construction of topological T-duality \cite{TopTdual1,TopTdual2} and briefly summarize the notion of Courant algebroid isomorphisms, which form the conceptual backbone of the duality. Section 4.3 then provides an explicit realization of T-duality in the spirit of the Gualtieri–Cavalcanti map \cite{guacav,GC}, where we transport a symplectic form on a torus bundle with symplectic fibers to its T-dual counterpart. After using the connection to decompose the space of invariant section of generalized tangent bundle we identified sections in $(\mathbb{X}, \Xi) \in (TM\oplus T^*M)/\mathbb{T}$ with a
quadruple $(X, f, \xi, g)$ as $\mathbb{X}=X+ f_ia^i$ and 
$\Xi = \xi +g^j\partial_j$ and defined a map 
$\varphi\colon (TM\oplus T^*M)/\mathbb{T}\to (T\hat{M}\oplus T^*\hat{M})/\mathbb{T}$. 
Following Equation (\ref{eq:TdualEquation}), we reinterpret T-duality in emergent gravity as the exchange of variables $f_i \leftrightarrow g^i$, which naturally induces a frame duality $\partial_i \to a^i$ in subsequent computations. This exchange can be viewed as an effective interchange between position and momentum coordinates. Ultimately, we show that under the GC map, a symplectic form is mapped to another symplectic form, related to the original one through the set of Equations (\ref{eq:Ap203a})–(\ref{eq:Ap205b}), which we have written explicitly for the case of emergent gravity with trivial H-flux.

\smallskip

Section 5, arguably the core of our work, unveils the deep interplay between emergent gravity and T-duality. Drawing inspiration from the seminal analysis of Seiberg and Witten \cite{ncft-sw}, which established a duality between semiclassical non-commutative and commutative descriptions of gauge theories, we demonstrate that in the context of emergent gravity on toroidal compactification, T-duality acts by turning background B-fields into 
$\theta$-transformations. In the absence of B-fields, the GC map leaves the Riemannian metric untouched; however, once B-fields are present, emergent gravity is naturally described by a two-step process - a B-transformation followed by a 
$\theta$-shift of the metric. On the T-dual side, intriguingly, this order is reversed, a fact elegantly encoded in the commutative diagram of Section 5.2.
Furthermore, we clarify that under the GC map a diffeomorphism maps to a diffeomorphism, hinting that the T-dual configuration may itself correspond to a distinct gravitational theory. The underlying mechanism is the Courant isomorphism, which realizes T-duality as an isomorphism between torus bundles (or equivalently, Courant algebroids), mapping the generalized metric to its inverse under the GC isomorphism. Once this conceptual framework is internalized, the subsequent derivations-from Equation (\ref{LHS-diagram}) to Equation (\ref{hatrelation})-follow immediately.

\smallskip

The final subsection, 5.3, extends our formalism to a completely general base manifold $B$. Remarkably, this generalization follows naturally from the procedure outlined in the preceding section. Beginning with the initial data (\ref{decomposition}) describing the emergent metric for a $\mathbb{T}^2$-fibration over an arbitrary base $B$, we derive explicit expressions for both the emergent metric components (Equation \ref{eq:EmergentMetric2}) and their T-dual counterparts (Equation \ref{eq:EmergentMetric2dual}). 

\smallskip

To summarize, in this work, we have formulated emergent gravity within the framework of generalized geometry and analyzed its behavior under T-duality. Starting from the emergent gravity data $(M, b, L, A, F, \theta, g, G, E_{+})$, we showed that the generalized metric $\mathcal{G}$, equivalently the subbundle $E_{+}=\mathrm{graph}(g+b)$, provides the natural geometric arena in which the symplectic structure $b$, the curvature $F$, and the emergent metric $g=g^{fl}(1+F\theta)$ unify. Within this framework, we derived explicit formulas for the T-dual of the symplectic form $b$ in the case of symplectic torus fibrations. These relations play the role of Buscher-type transformation laws for the symplectic sector, thereby extending duality beyond the metric alone. Moreover, emergent gravity was reinterpreted as a composition of a $B$-transform and a $\theta$-transform acting on a flat background, from which the Seiberg–Witten relations follow naturally as generalized geometric identities.
On the dual side $(\widehat{M}, \widehat{\mathcal{G}}, \widehat{E}_{+})$, we established the functorial transport of the generalized metric under T-duality via the GC map. In the fluxless case $H=db=0$, the correspondence space construction with the canonical two-form $K$ reproduces the standard relation $p^{*}H-\widehat{p}^{*}\widehat{H}=dK$, while simultaneously encoding the emergent gravity data. In addition, we obtained a genuine generalization of the Buscher rules to metrics on $\mathbb{T}^2$-fibrations, extending the usual $\mathbb{T}^1$-fiber formulation and matching recent results \cite{Nikolaus-Waldorf} in the literature from an independent generalized geometric perspective. Taken together, these results demonstrate that emergent gravity admits a natural and structurally coherent formulation within generalized geometry, and that its T-dual can be described in the same language. The explicit transformation laws for the symplectic form and generalized metric provide a concrete step toward a unified treatment of emergent gravity and topological T-duality, and lay the groundwork for future investigations in the presence of nontrivial flux.

\smallskip

The case with non-trivial 
H-flux presents subtleties, as one encounters H-twisted symplectic and H-twisted Poisson structures. In the former case, the 2-form is no longer closed; instead, its exterior derivative satisfies 
$(H=db)$. Consequently, the structure ceases to be symplectic and no longer induces a standard Poisson bracket. Conversely, the corresponding Poisson structure fails to satisfy the Jacobi identity, holding only up to an exact term proportional to $dH$. In the presence of H-flux, there is thus no genuine symplectic structure, and under T-duality, one cannot simply transform it into a conventional Poisson structure. This complication does not, however, obstruct the generalized geometry framework, where one can still define H-twisted objects using the GC map \cite{GC}. After the twist, however, the isotropic integrable subbundles with respect to the twisted Dorfman bracket (\ref{twisted dorfman}) fail to remain integrable, so that the graphs (\ref{graphs}) no longer represent a generalized metric. Considering T-duality for strings or Dp-branes propagating in curved backgrounds with non-trivial 
H-flux leads naturally to situations where the effective description involves so-called non-geometric fluxes \cite{nongeometry,nongeometry2,nongeometry3,nongeometry4,nongeometry5, nongeometry6, nongeometry7, nongeometry8}. 
It would be interesting to unravel the full geometry behind these non-geometric fluxes particularly in the broader context of emergent gravity.

\smallskip

In the context of emergent gravity, we have a framework \cite{hsy-emergent, hsy-emergent2, lee1, lee2, lee3} where one begins with a noncommutative 
$U(1)$ gauge theory, which is Seiberg–Witten equivalent to a commutative $U(1)$ gauge theory with a dynamically induced metric. The commutative description is characterized by a line bundle, its curvature, and an underlying Poisson structure, which together provide the initial data for the emergent geometry. However, in the presence of a nontrivial H-flux, these data are only defined up to an H-twist, and a systematic generalization is required to accommodate the flux. This raises several open questions: What is the appropriate notion of gauge theory on a space with H-flux? How should we define a Poisson structure in such a background? And in what way are these generalized structures related so as to give rise to a gravitational theory? In a recent work 
\cite{RS} dealing with the completely general case on a nilmanifold with B-fields that are symplectic both on the base and the fiber, we have furnished examples of the T-duality of emergent gravity on self-T-dual nilmanifolds with zero flux. For non-zero flux, only in some specific scenario it was shown that such a duality of emergent gravity fails to exist. In particular with H-flux, a non-commutative line bundle must be replaced by a $U(1)$ gerbe, since for $H=0$ one has a trivial gerbe equivalent to an ordinary line bundle, whereas a nontrivial 
H-flux requires a genuinely twisted object. The resulting 
H-twisted Poisson structure should lead to a non-commutative $U(1)$ gerbe, and one may attempt to formulate an action functional for this theory and study its field-theoretic content. Such constructions are naturally phrased in the language of higher gauge theories or functorial field theories. Locally, these noncommutative gerbes may admit a Seiberg–Witten-like map relating them to commutative data.

\smallskip

Furthermore, the work of Hull et al. \cite{Hull-gerbe, Hull-gerbe2} demonstrates how holomorphic line bundles and Kähler geometry lift to bi-holomorphic line bundles and generalized complex geometry under T-duality, suggesting a path toward formulating emergent gravity in the gerby setting. Related approaches using non-Abelian gerbes as higher geometric analogues of principal bundles have been developed in \cite{gerbe-CMP}. Finally, recent progress on topological T-duality in gerbe-theoretic language \cite{Nikolaus-Waldorf} appears particularly promising for developing a precise formulation of emergent gravity in twisted backgrounds, and we plan to return to these questions in future work.

\smallskip

In the presence of a nontrivial $H=db$, the construction of emergent gravity remains feasible. A particularly interesting case arises when $H=b \wedge l$, $l$ being the so-called
Lee form \cite{lee},
leading to a locally conformal symplectic (LCS) manifold \cite{LCS, LCS2}. As emphasized in Appendix A of \cite{hsy-app}, the key property of an LCS manifold is that its local structure is indistinguishable from that of a symplectic manifold, allowing the use of local Darboux charts. We believe that LCS manifolds provide a natural setting to describe phenomena such as cosmic inflation and black hole geometries within the emergent gravity framework. In this scenario, the B-fields induce a diffeomorphism symmetry that is strictly richer than the ordinary symplectic case. This strongly suggests that the most general diffeomorphisms should be generated by generic 
B-fields with arbitrary 
$H=db$, although a fully general construction of emergent gravity in this setting remains an open problem.

\smallskip

Finally, while the emergent gravity paradigm is widely believed to extend beyond Kähler manifolds, a systematic formulation for homogeneous non-Kähler manifolds is still lacking. Recent progress, however, has been made in \cite{Tnil}, where the authors developed an infinitesimal version of T-duality on homogeneous compact manifolds with a natural torus bundle structure, using a Lie algebraic approach. Remarkably, their framework allows, under suitable conditions, the construction of T-duals for nilmanifolds with nontrivial 
H-flux. They exploit the Gualtieri–Cavalcanti formalism \cite{guacav} to transport generalized complex branes-submanifolds generalizing holomorphic and coisotropic branes-in a manifestly invariant way. The emergent gravity framework has also been extended to homogeneous compact manifolds associated with nilpotent Lie groups, an intriguing direction for further research that has been investigated by one of the authors in \cite{RS}.

\bigskip

{\bf Acknowledgements}  
One of the authors RR would like to thank Hyun Seok Yang, Giovanni Landi, Paolo Antonioni, Leonardo Soriani, Leandro Gomes and Victor Rivelles for many useful discussions on related topics during different stages of the project. 
RR would also like to thank IMPA at Rio de Janeiro and the organizers of the conference
on string geometries and dualities where this project was first conceived.
The research of RR was partly supported by PNPD-CAPES Matemática  (32003013012P1)
with the grant 88887.356891/2019-00.
RR is grateful to UPES, Dehradun and University of Delhi for the hospitality and support. RR would also like to acknowledge his debt to the people of India for their steady support of the study of basic science.

\bigskip

{\bf Conﬂict of Interest}
The authors declare no conﬂict of interest.

\bigskip

{\bf Data Availability Statement}
Data sharing not applicable to this article as no datasets were generated or analysed during the current study.

\bigskip

{\bf Keywords}
Topological T-duality, Generalized geometry, H flux, Courant algebroids, Emergent gravity

\newpage

\appendix 

\section{T-dual Generalized Metric for a 2-Torus Fibration} 
In this appendix we give a detailed derivation of the T-dual of an arbitrary generalized metric on a 
$\mathbb{T}^2$-fibration. We will consider a spacetime $M$ which is a $\mathbb{T}^2$-fibration over an arbitrary base manifold $B$, to find the T-dual generalized metric we first decompose, according to the direct sum $T^*M\cong T^*B\oplus \langle a_1\rangle\oplus \langle a_2\rangle $ and $TM\cong TB\oplus \langle \partial_1\rangle\oplus \langle \partial_2\rangle $ given by the connection $a=(a_1,a_2)$, the background metric $g$ and the 2-form $b$

\bea 
\begin{aligned}
\label{eq:Ap00}
g&=g_{0,11} a_1\odot a_1 + g_{0,22}a_2\odot a_2+g_{0,12} a_1\odot a_2 +g_{1,1}\odot a_1+g_{1,2}\odot a_2+g_2,\\
b&=b_{0,12}a_1\wedge a_2 + b_{1,1}\wedge a_1+b_{1,2}\wedge a_2+b_2.
\end{aligned}
\eea 
To ease notation we will write $b_0$ instead of $b_{0,12}$. The generalized metric is characterized by the graph of $(g+b)$ denoted by
$E_+$ (See Section 3). In the mentioned decomposition of $TM$, described by
the connection a general element of the tangent bundle is given by
$X+f_1\partial_1+f_2\partial_2$. Thus the graph of $(g+b)$ is given by 
\be\begin{aligned}
\label{eq:Ap01}
X&+f_1 {\partial}_{1}\\
&+ {f}_2 {\partial}_{2}\\&+ \iota_X {g}_2+{g_{1,1}}{f}_1+{g_{1,2}}{f}_2+\iota_X {b}_2-{b_{1,1}}{f}_1-{b_{1,2}}{f}_2\\&
+\left[{g_{0,11}}{f}_1 + {g_{0,12}}{f
}_2 + {g_{1,1}}(X)-{b}_0{f}_2+{b_{1,1}}(X)\right] {a}_{1}\\&
+\left[{g_{0,22}}{f}_2 + {g_{0,12}}{f
}_1 + {g_{1,2}}(X)+{b}_0{f_1}+{b_{1,2}}(X)\right] a_{2}.\end{aligned}\ee
Under the Courant algebroid isomorphism implementing T-duality such elements of the graph are mapped to elements of $\varphi(E_+)$ of the form:
\be
\begin{aligned}
\label{eq:Ap02}
X&+\left[g_{0,11} f_{1}+g_{0,12} f_{2}+g_{1,1}(X)-b_{0} f_{2}+b_{1,1}(X)\right] \tilde{\partial}_{1}\\
&+ \left[g_{0,22} f_{2}+g_{0,12} f_{1}+g_{1,2}(X)+b_{0} f_{1}+b_{1,2}(X)\right] \tilde{\partial}_{2}\\&+ \iota_Xg_2+g_{1,1}f_1+g_{1,2}f_2+\iota_xb_2-b_{1,1}f_1-b_{1,2}f_2\\&
+f_{1} \tilde{a}_{1}\\&
+f_{2} \tilde{a}_{2}.
\end{aligned}
\ee
We seek a generalized metric $(\tilde{g}+\tilde{b})$ such that the graph $\tilde{C}_+$ correspond to the elements defined by $\varphi(E_+)$, i.e. elements of the form of Equation (\ref{eq:Ap02}). To proceed we decompose $(\tilde{g}+\tilde{b})$ with respect to the splitting $T^*M=T^*B\oplus \langle a_1\rangle\oplus \langle a_2\rangle$. With such a decomposition a general element of the graph, of the unknown, $(\tilde{g}+\tilde{b})$ can be written as:
\be\begin{aligned}
\label{eq:Ap03}
X&+\tilde{f}_1 \tilde{\partial}_{1}\\
&+ \tilde{f}_2 \tilde{\partial}_{2}\\&+ \iota_X \tilde{g}_2+\widetilde{g_{1,1}}\tilde{f}_1+\widetilde{g_{1,2}}\tilde{f}_2+\iota_X \tilde{b}_2-\widetilde{b_{1,1}}\tilde{f}_1-\widetilde{b_{1,2}}\tilde{f}_2\\&
+\left[\widetilde{g_{0,11}}\tilde{f}_1 + \widetilde{g_{0,12}}\tilde{f
}_2 + \widetilde{g_{1,1}}(X)-\tilde{b}_0\tilde{f}_2+\widetilde{b_{1,1}}(X)\right] \tilde{a}_{1}\\&
+\left[\widetilde{g_{0,22}}\tilde{f}_2 + \widetilde{g_{0,12}}\tilde{f
}_1 + \widetilde{g_{1,2}}(X)+\tilde{b}_0\tilde{f_1}+\widetilde{b_{1,2}}(X)\right] \tilde{a}_{2}.\end{aligned}\ee

\smallskip

Taking  $\tilde{f_1}=g_{0,11} f_{1}+g_{0,12} f_{2}+g_{1,1}(X)-b_{0} f_{2}+b_{1,1}(X)$ and $\tilde{f_2}=g_{0,22} f_{2}+g_{0,12} f_{1}+g_{1,2}(X)+b_{0} f_{1}+b_{1,2}(X)$, the $\partial_1$ and $\partial_2$ components agree (notice that $X+\tilde{f}_1\partial_1+\tilde{f}_2\partial_2$ is still a general element of $TM$). The condition for $\tilde{C}_+$ to be the graph of $(\tilde{g}+\tilde{b})$ implies an equality for the expressions of (\ref{eq:Ap02}) and (\ref{eq:Ap03}), which in turn yields a system of equations that can be used to find an explicit description of $(\tilde{g}+\tilde{b})$.

\smallskip

The equations arising from the $\tilde{a}_1$ components and coefficients of $f_1$ and $f_2$, respectively, are:

$$1=\widetilde{g_{0,11}}\left[g_{0,11}\right]+\widetilde{g_{0,12}}\left[g_{0,12}+b_{0}\right]-\widetilde{b_{0}}\left[g_{0,12}+b_{0}\right],
$$
$$0=\widetilde{g_{0,11}}\left[g_{0,12}-b_0\right]+\widetilde{g_{0,12}}\left[g_{0,22}\right]-\widetilde{b_{0}}\left[g_{0,22}\right].$$
Using the previous equations we get:

\be
\label{eq:Ap04}
\widetilde{g_{0,11}}=\frac{-g_{0,22}}{g_{0,12}^{2}-b_{0}^{2}-g_{0,11} g_{0,22} },\qquad \widetilde{g_{0,22}}=\frac{-g_{0,11}}{g_{0,12}^{2}-b_{0}^{2}-g_{0,11} g_{0,22} },
\ee
and 
\be
\label{eq:Ap05}
\widetilde{g_{0,12}}-\tilde{b}_{0}=\frac{g_{0,12}-b_{0}}{g_{0,12}^{2}-b_{0}^{2}-g_{0,11} g_{0,22}}.  
\ee

On the other hand,  the equation arising from the $\tilde{a}_2$ components and coefficients of $f_1$ and $f_2$, respectively, are:

$$0=\widetilde{g_{0,22}}\left[g_{0, 12}+b_0\right]+(\widetilde{g_{0,12}}+\tilde{b_{0}})\left[g_{0,11}\right],$$

$$1=\widetilde{g_{0,22}}\left[g_{0, 22}\right]+(\widetilde{g_{0,12}}+\tilde{b_{0}})\left[g_{0,12}-b_{0}\right].$$
Using Equation (\ref{eq:Ap04}) and the previous equations, one obtains:

\be
\label{eq:R1}
\widetilde{g_{0,12}}=\frac{g_{0,12}}{g_{0,12}^{2}-b_{0}^{2}-g_{0,11} g_{0,22}},\qquad \tilde{b}_{0}=\frac{b_{0}}{g_{0,12}^{2}-b_{0}^{2}-g_{0,11} g_{0,22}}.
\ee
Now, the 1-form contracted with $X$ from the $a_1$ and $a_2$ components are, as follows:  
$$0 = \widetilde{g_{0,11}}\left[g_{1,1}(X)+b_{1,1}(X) \right]+\widetilde{g_{0,12}}\left[g_{1,2}(X)+b_{1,2}(X) \right]+\widetilde{g_{1,1}}(X)-\tilde{b_0}\left[g_{1,2}(X)+b_{1,2}(X) \right]+\widetilde{b_{1,1}}(X),$$
and
$$0 = \widetilde{g_{0,22}}\left[g_{1,2}(X)+b_{1,2}(X) \right]+\widetilde{g_{0,12}}\left[g_{1,1}(X)+b_{1,1}(X) \right]+\widetilde{g_{1,2}}(X)+\tilde{b_0}\left[g_{1,1}(X)+b_{1,1}(X) \right]+\widetilde{b_{1,2}}(X).$$
Using (\ref{eq:Ap04}), we get

\be
\label{eq:Ap0505}
\widetilde{g_{1,1}}(X)+\widetilde{b_{1,1}}(X)= \frac{g_{0,22}(g_{1,1}+b_{1,1})-(g_{0,12}-b_0)(g_{1,2}+b_{1,2})}{g_{0,12}^{2}-b_{0}^{2}-g_{0,11} g_{0,22}}.
\ee

Furthermore, the $f_1$ and $f_2$ coefficients from the $T^*B$ components read as:
$$g_{1,2}-b_{1,2}=\left[\widetilde{g_{1,1}}-\widetilde{b_{1,1}}\right]\left[g_{0,12}-b_0 \right]+\left[\widetilde{g_{1,2}}-\widetilde{b_{1,2}}\right]\left[g_{0,22} \right],$$
and
$$g_{1,1}-b_{1,1}=\left[\widetilde{g_{1,1}}-\widetilde{b_{1,1}}\right]\left[g_{0,11} \right]+\left[\widetilde{g_{1,2}}-\widetilde{b_{1,2}}\right]\left[g_{0,12}+b_0 \right].$$
From the previous equations one obtains: 
\be
\label{eq:Ap06}
\widetilde{g_{1,1}}(X)-\widetilde{b_{1,1}}(X)= \frac{(g_{0,12}+b_{0})(g_{1,2}-b_{1,2})-g_{0,22}(g_{1,1}-b_{1,1})}{g_{0,12}^{2}-b_{0}^{2}-g_{0,11} g_{0,22}},
\ee
and 
\be
\label{eq:Ap07}
\widetilde{g_{1,2}}(X)-\widetilde{b_{1,2}}(X)= \frac{(g_{0,12}-b_{0})(g_{1,1}-b_{1,1})-g_{0,11}(g_{1,2}-b_{1,2})}{g_{0,12}^{2}-b_{0}^{2}-g_{0,11} g_{0,22}}.
\ee
Using (\ref{eq:Ap0505}), (\ref{eq:Ap06}) and (\ref{eq:Ap07}), we obtain:

\be
\label{eq:ApR2}
\begin{aligned}
\widetilde{g_{1,1}}(X)=&\left[ \frac{-g_{0,12}}{g_{0,12}^{2}-b_{0}^{2}-g_{0,11} g_{0,22}} b_{1,2}(X) \right.\\&+\frac{b_{0}}{g_{0,12}^{2}-b_{0}^{2}-g_{0,11} g_{0,22}} g_{1,2}(X)\\&\left. + \frac{g_{0,22}}{g_{0,12}^{2}-b_{0}^{2}-g_{0,11} g_{0,22}} b_{1,1}(X)\right], 
\end{aligned}
\ee
and 
\be
\label{eq:ApR3}
\begin{aligned}
\widetilde{b_{1,1}}(X)=&\left[ \frac{-g_{0,12}}{g_{0,12}^{2}-b_{0}^{2}-g_{0,11} g_{0,22}} g_{1,2}(X)\right.\\&+\frac{b_{0}}{g_{0,12}^{2}-b_{0}^{2}-g_{0,11} g_{0,22}} b_{1,2}(X)\\ &\left.+ \frac{g_{0,22}}{g_{0,12}^{2}-b_{0}^{2}-g_{0,11} g_{0,22}} g_{1,1}(X)\right].
\end{aligned}
\ee

Similarly we may obtain:
\be
\label{eq:R4}
\begin{aligned}
\widetilde{g_{1,2}}(X)=&\left[ \frac{-g_{0,12}}{g_{0,12}^{2}-b_{0}^{2}-g_{0,11} g_{0,22}} b_{1,1}(X) \right.\\&-\frac{b_{0}}{g_{0,12}^{2}-b_{0}^{2}-g_{0,11} g_{0,22}} g_{1,1}(X)\\&\left. + \frac{g_{0,11}}{g_{0,12}^{2}-b_{0}^{2}-g_{0,11} g_{0,22}} b_{1,2}(X)\right], 
\end{aligned}
\ee
and 
\be
\label{eq:R5}
\begin{aligned}
\widetilde{b_{1,2}}(X)=&\left[ \frac{-g_{0,12}}{g_{0,12}^{2}-b_{0}^{2}-g_{0,11} g_{0,22}} g_{1,1}(X)\right.\\&-\frac{b_{0}}{g_{0,12}^{2}-b_{0}^{2}-g_{0,11} g_{0,22}} b_{1,1}(X)\\ &\left.+ \frac{g_{0,11}}{g_{0,12}^{2}-b_{0}^{2}-g_{0,11} g_{0,22}} g_{1,2}(X)\right].
\end{aligned}
\ee

\bigskip 
At last the equation obtained from the form contracted with $X$ from the $T^*B$ component is:
$$\iota_X[g_2+b_2]=\iota_X(\tilde{g}_2+\tilde{b}_2)+\left[\widetilde{g_{1,1}}-\widetilde{b_{1,1}} \right]\iota_X\left[g_{1,1}+b_{1,1}\right]+\left[\widetilde{g_{1,2}}-\widetilde{b_{1,2}} \right]\iota_X\left[g_{1,2}+b_{1,2}\right]. $$
Using (\ref{eq:Ap04}) - (\ref{eq:Ap07})
and distinguishing the symmetric and anti-symmetric components of
$(\tilde{g}+\tilde{b})$ we arrive at:
\be
\label{eq:R6}
\begin{aligned}\tilde{g}_2=g_2+&\left[\frac{b_0}{g_{0,12}^{2}-b_{0}^{2}-g_{0,11} g_{0,22}}\right]\left[ g_{1,1}\odot b_{1,2}- g_{1,2}\odot b_{1,1}\right]\\
&-\left[\frac{g_{0,12}}{g_{0,12}^{2}-b_{0}^{2}-g_{0,11} g_{0,22}}\right]\left[ g_{1,1}\odot g_{1,2}- b_{1,1}\odot b_{1,2}\right]\\+&\left[\frac{g_{0,11}}{g_{0,12}^{2}-b_{0}^{2}-g_{0,11} g_{0,22}}\right]\left[ g_{1,2}\odot g_{1,2}- b_{1,2}\odot b_{1,2}\right]\\+&\left[\frac{g_{0,22}}{g_{0,12}^{2}-b_{0}^{2}-g_{0,11} g_{0,22}}\right]\left[ g_{1,1}\odot g_{1,1}- b_{1,1}\odot b_{1,1}\right],\end{aligned}
\ee
\be
\label{eq:R7}
\begin{aligned}\tilde{b}_2=b_2+&\left[\frac{b_0}{g_{0,12}^{2}-b_{0}^{2}-g_{0,11} g_{0,22}}\right]\left[ g_{1,1}\wedge g_{1,2}-b_{1,1}\wedge b_{1,2}\right]\\
&-\left[\frac{g_{0,12}}{g_{0,12}^{2}-b_{0}^{2}-g_{0,11} g_{0,22}}\right]\left[ g_{1,1}\wedge b_{1,2}+g_{1,2}\wedge b_{1,1}\right]\\
+&\left[\frac{g_{0,11}}{g_{0,12}^{2}-b_{0}^{2}-g_{0,11} g_{0,22}}\right]\left[ g_{1,2}\wedge b_{1,2}\right]\\+&\left[\frac{g_{0,22}}{g_{0,12}^{2}-b_{0}^{2}-g_{0,11} g_{0,22}}\right]\left[ g_{1,1}\wedge b_{1,1}\right].\end{aligned}
\ee

\bigskip

At last we summarize the results so far derived in the appendix. The dual generalized emergent metric $(\tilde{g}+\tilde{b})$ has a decomposition given by 
\bea 
\begin{aligned}
\label{eq:ApD}
\tilde{g}&=\tilde{g}_{0,11} a_1\odot a_1 + \tilde{g}_{0,22}a_2\odot a_2+\tilde{g}_{0,12} a_1\odot a_2 +\tilde{g}_{1,1}\odot a_1+\tilde{g}_{1,2}\odot a_2+\tilde{g}_2,\\
\tilde{b}&=\tilde{b}_{0,12}a_1\wedge a_2 + \tilde{b}_{1,1}\wedge a_1+\tilde{b}_{1,2}\wedge a_2+\tilde{b}_2,
\end{aligned}
\eea 
where defining $k=g_{0,12}^{2}-b_{0}^{2}-g_{0,11} g_{0,22}$ the components are given by the following equations:
\be
\label{eq:RF1}
\widetilde{g_{0,12}}=\frac{g_{0,12}}{k},\qquad \tilde{b}_{0}=\frac{b_{0}}{k}, \qquad
\widetilde{g_{0,11}}=\frac{-g_{0,22}}{k},\qquad \widetilde{g_{0,22}}=\frac{-g_{0,11}}{k},
\ee

\be
\label{eq:RF4}
\begin{aligned}
\widetilde{g_{1,1}}(X)=\frac{1}{k}&\left[-g_{0,12} b_{1,2}(X)+b_{0} g_{1,2}(X) + g_{0,22} b_{1,1}(X)\right], 
\end{aligned}
\ee

\be
\label{eq:RF5}
\begin{aligned}
\widetilde{g_{1,2}}(X)=\frac{1}{k}&\left[-g_{0,12}b_{1,1}(X)-b_{0} g_{1,1}(X) + g_{0,11} b_{1,2}(X)\right], 
\end{aligned}
\ee

\be
\label{eq:RF6}
\begin{aligned}
\widetilde{b_{1,1}}(X)=\frac{1}{k}&\left[-g_{0,12} g_{1,2}(X)+b_{0} b_{1,2}(X)+ g_{0,22} g_{1,1}(X)\right],
\end{aligned}
\ee

\be
\label{eq:RF7}
\widetilde{b_{1,2}}(X)=\frac{1}{k}\left[ -g_{0,12} g_{1,1}(X)-b_{0} b_{1,1}(X)+ g_{0,11} g_{1,2}(X)\right],
\ee

\be
\label{eq:RF8}
\begin{aligned}\tilde{g}_2=g_2+&\frac{1}{k} \left[b_0(g_{1,1}\odot b_{1,2}- g_{1,2}\odot b_{1,1})
-g_{0,12}(g_{1,1}\odot g_{1,2}- b_{1,1}\odot b_{1,2})\right.\\&\left.+g_{0,11}( g_{1,2}\odot g_{1,2}- b_{1,2}\odot b_{1,2})+g_{0,22}( g_{1,1}\odot g_{1,1}- b_{1,1}\odot b_{1,1})\right],\end{aligned}
\ee

\be
\label{eq:RF9}
\begin{aligned}\tilde{b}_2=b_2+&\frac{1}{k}\left[b_0( g_{1,1}\wedge g_{1,2}-b_{1,1}\wedge b_{1,2})-g_{0,12}( g_{1,1}\wedge b_{1,2}+g_{1,2}\wedge b_{1,1})\right.\\
+&\left.g_{0,11}( g_{1,2}\wedge b_{1,2})+g_{0,22}( g_{1,1}\wedge b_{1,1})\right].\end{aligned}
\ee

These expressions describe the most general form of the T-dual metric and B-field for a $\mathbb{T}^2$-fibration. Consequently, the relations given in Eqs. (A.17)–(A.23) may be regarded as a natural generalization of the classical Buscher rules (\cite{Buscher}) to the case of an arbitrary metric on a $\mathbb{T}^2$-fibered background.

\bigskip

In the special case where the metric is diagonal along the fiber directions, and the B-field has no component entirely supported on the fibers, i.e. $g_{0,12}=0$ and $b_{0}=0$, these expressions simplify considerably and assume a form much closer to the standard Buscher transformations. Furthermore, in the degenerate limit in which one of the two circle fibers is removed, reducing the $\mathbb{T}^2$-fibration to a single $S^1$-fibration, the above formulas reduce precisely to the usual Buscher rules for Abelian T-duality.

\be
\label{eq:RF2}
\widetilde{g_{0,11}}=\frac{1}{g_{0,11}},\qquad \widetilde{g_{0,22}}=\frac{1}{g_{0,22}},
\ee

\be
\begin{aligned}
\widetilde{g_{1,1}}(X)=\frac{b_{1,1}(X)}{g_{0,11}} , \qquad
\widetilde{g_{1,2}}(X)=\frac{b_{1,2}(X)}{g_{0,22}}, 
\end{aligned}
\ee

\be
\begin{aligned}
\widetilde{b_{1,1}}(X)=\frac{g_{1,1}(X)}{g_{0,11}},\qquad
\widetilde{b_{1,2}}(X)=\frac{  g_{1,2}(X)}{g_{0,22}},
\end{aligned}
\ee

\be
\begin{aligned}\tilde{g}_2=g_2
-\frac{ g_{1,2}\odot g_{1,2}- b_{1,2}\odot b_{1,2}}{g_{0,22}}-\frac{g_{1,1}\odot g_{1,1}- b_{1,1}\odot b_{1,1}}{g_{0,11}} ,\end{aligned}
\ee

\be
\begin{aligned}\tilde{b}_2=b_2+ \frac{ g_{1,2}\wedge b_{1,2}}{g_{0,22}} + \frac{ g_{1,1}\wedge b_{1,1}}{g_{0,11}}.\end{aligned}
\ee

\bibliographystyle{unsrt}
\bibliography{Bibliography.bib}

\end{document}